\documentclass[preprint2]{proto}
\usepackage{times}

\def\lb#1{{\protect\linebreak[#1]}}
\def\httpyarko{{http://\lb{2}www.youtube.\lb{2}com/\lb{2}watch?v=\lb{2}kzlgxqXtxYs}}
\def\httpmira{{http://\lb{2}sirrah.\lb{2}troja.mff.\lb{2}cuni.cz/{\~{ }}\lb{2}mira/\lb{2}mp/\lb{2}phdth/}}
\def\httpearn{{http://\lb{2}earn.dlr.\lb{2}.de/\lb{2}nea}}
\def\httpbin{{http://\lb{2}www.asu.\lb{2}cas.cz/{\~{ }}\lb{2}asteroid/\lb{2}binastdata.\lb{2}htm}}
\newcommand{\refs}{\par\noindent\hangindent=1pc\hangafter=1}
\begin{document}

\title{\textbf{\LARGE The Yarkovsky and YORP Effects}}

\author {\textbf{\large David Vokrouhlick\'y}}
\affil{\small\em Institute of Astronomy, Charles University, Prague}

\author {\textbf{\large William F. Bottke}}
\affil{\small\em Department of Space Studies,
                 Southwest Research Institute, Boulder}

\author {\textbf{\large Steven R. Chesley}}
\affil{\small\em Jet Propulsion Laboratory, California Institute of Technology,
                 Pasadena}

\author {\textbf{\large Daniel J. Scheeres}}
\affil{\small\em Aerospace Engineering Sciences, University of Colorado, Boulder}

\author {\textbf{\large Thomas S. Statler}}
\affil{\small\em Astrophysical Institute, Ohio University, Athens and \\
                 Department of Astronomy, University of Maryland, College Park}

\begin{abstract}
\begin{list}{ }{\rightmargin 0.3in}
\baselineskip = 11pt
\parindent=1pc
{\small The Yarkovsky effect describes a small but significant force that affects
 the orbital motion of meteoroids and asteroids smaller than $30-40$ kilometers in diameter.
 It is caused by sunlight; when these bodies heat up in the Sun, they eventually re-radiate
 the energy away in the thermal waveband, which in turn creates a tiny thrust. This recoil
 acceleration is much weaker than solar and planetary gravitational forces, but it can
 produce measurable orbital changes over decades and substantial orbital effects over millions
 to billions of years. The same physical phenomenon also creates a thermal torque that,
 complemented by a torque produced by scattered sunlight, can modify the rotation rates
 and obliquities of small bodies as well. This rotational variant has been coined the
 Yarkovsky-O'Keefe-Radzievskii-Paddack (YORP) effect. During the past decade or so, the
 Yarkovsky and YORP effects have been used to explore and potentially resolve a number
 of unsolved mysteries in planetary science dealing with small bodies. 
 Here we review the main results to date, and preview the goals for future work.
 \\~\\~\\~} 
 
\end{list}
\end{abstract}  

\section{\textbf{INTRODUCTION}}
Interesting problems in science usually have a long and complex history. It 
is rare, though, that they have a prehistory or perhaps even mythology. Yet, until
recently this was the case of the Yarkovsky effect. Ivan O. Yarkovsky, a 
Russian civil engineer born in a family of Polish descent, noted in a
privately-published pamphlet ({\em Yarkovsky}, 1901; {\em Beekman}, 2006) 
that heating a prograde-rotating planet should produce a transverse 
acceleration in its motion and thus help to counter-balance the assumed
drag from the then-popular ether hypothesis. While this context of Yarkovsky's 
work was mistaken and he was only roughly able to estimate the magnitude 
of the effect, he succeeded in planting the seed of an idea that a century 
later blossomed into a full-fledged theory of how the orbits of small objects 
revolving about the Sun are modified by the absorption and re-emission 
of solar energy. 

It is mainly Ernst J. \"Opik who is to be credited for keeping 
Yarkovsky's work alive and introducing it to western literature, long after 
the original pamphlet had been lost ({\em \"Opik}, 1951). Curiously, at about the
same time, similar ideas also started to appear in Russian regular scientific
literature through the works of Vladimir V. Radzievskii and his
collaborators ({\em Radzievskii}, 1952b). While Radzievskii was also the first
to consider the effects of systematic photon thrust on a
body's rotation, his concept was based on a variable albedo coefficient across
the surface ({\em Radzievskii}, 1952a). However, there is no strong evidence of 
large enough albedo variations over surfaces of asteroids or meteoroids.
Stephen J. Paddack and John O'Keefe pushed the idea forward by realizing that
irregular shape, and thermal radiation rather than just the reflected
sunlight, will more efficiently change the meteoroid's spin rate. Thence,
the Yarkovsky-O'Keefe-Radzievskii-Paddack effect (YORP for short) was born
as an alter ego of the Yarkovsky effect little more than half a century after 
Yarkovsky's work (see {\em Paddack}, 1969; {\em Paddack and Rhee}, 1975; 
and {\em Rubincam}, 2000 for summing up the history and coining the terminology).
Radzievskii's school also briefly touched upon a concept of a radiation-induced
acceleration of synchronous planetary satellites ({\em Vinogradova and Radzievskii},
1965), an idea that much later re-appeared in a slightly different form as 
a binary YORP (or BYORP for short) effect ({\em \'Cuk and Burns}, 2005). 

The three decades from the 1950's to the 1970's saw a slow revival and emergence of 
the concepts that eventually resulted in today's Yarkovsky and YORP effects.
The works that led to a major resurgence in these studies, however, occurred 
in the second half of the 1990's through the work
of David P. Rubincam and Paolo Farinella. Interestingly, both were studying
thermal perturbations of artificial satellite motion. With that expertise
they realized a direct link between the orbital effects acting on the geodynamics
artificial satellites such as LAGEOS and the orbital effects on small meteoroids
(e.g., {\em Afonso et~al.}, 1995; {\em Rubincam}, 1995, 1998; {\em Farinella et~al.}, 
1998). 

From there, 
a momentum was gained and a wealth of new results appeared, with applications
extending to a dynamics of small asteroids and their populations (e.g., {\em Bottke 
et~al.}, 2002a, 2006). Studies of the Yarkovsky effect were soon followed by
those of the YORP effect ({\em Rubincam}, 2000). Today, both effects belong to
a core culture in planetary sciences, as well as beyond (e.g., \httpyarko), and
have become an important part in the agenda of space missions (e.g., {\em 
Lauretta et~al.}, 2015). Especially after the spectacular discovery of 
the ``once lost'' Yarkovsky pamphlet in Russian archives by Dutch amateur 
astronomer George Beekman (see {\em Beekman}, 2006), it seems timely to review 
the current knowledge of the Yarkovsky and YORP effects. This
effort could start with a translation, and perhaps a commented edition, of the 
Yarkovsky work (presently available in the original form as an Appendix of Miroslav 
Bro\v{z}'s Thesis, \httpmira). We look forward to future historians editing the 
more than a century long story of the Yarkovsky and YORP effects, with all the 
known and possibly hidden roots, into a consolidated picture.

Leaving historical issues to their own life, we now turn to current scientific
issues related to the Yarkovsky and YORP effects. There are several good technical 
reviews already existing in the literature (e.g., {\em Bottke et~al.}, 2002a, 2006). While
not always possible, we try to avoid discussing the same topics as presented in 
these previous texts. For instance, we do not review the elementary 
concepts of the Yarkovsky and YORP effects, assuming the reader is familiar with
them. Rather, we try to focus on new results and ideas that emerged during the past 
decade and that will lead to research efforts in the next several years.

\bigskip

\begin{center}
 \textbf{2. THEORY OF THE YARKOVSKY \\ AND YORP EFFECTS}
\end{center}

\bigskip We start with the simplest analytical models of the Yarkovsky and YORP
effects (Sec.~2.1). This is because they provide useful insights, such as scalings 
with several key parameters, and their results are correct to leading order. 
They also allow us to understand why modeling of the YORP effect is inevitably more
complicated than modeling of the Yarkovsky effect. And yet, the quality
of the Yarkovsky and YORP effects detections, as well as other applications, have
reached a level that requires more accurate models to be used. First steps towards
these new models have been taken recently and these are briefly reviewed
in Sec.~2.2.

\bigskip
\noindent
 \textbf{2.1 Classical models}
\bigskip

\noindent{\bf The Yarkovsky effect.-- }Absorbed and directly reflected sunlight 
does not tend to produce long-term dynamical effects as far as orbital motion is 
concerned (e.g., {\em Vokrouhlick\'y and Milani}, 2000; {\em \v{Z}i\v{z}ka and
Vokrouhlick\'y}, 2011). The Yarkovsky effect thus fundamentally depends on emitted 
thermal radiation and requires a body to have a non-zero thermal inertia. Any meaningful 
evaluation of the Yarkovsky effect, therefore, requires a thermophysical model of 
that body. Fortunately, an evaluation of the Yarkovsky effect imposes a minimum of 
requirements on the shape of the body; even a simple spherical model provides us 
with a fair approximation of how the body will orbitally evolve. 

While the Yarkovsky effect results in variations to all of the orbital elements, 
what is distinct from most other perturbations is the secular effect in
the semimajor axis $a$. Therefore, we only discuss this contribution.
Assuming (i) a linearization of the surface boundary condition, (ii) a rotation 
about a spin axis fixed in the inertial space (at least on a timescale comparable
with the revolution about the Sun), and (iii) a circular orbit about the Sun,
one easily finds that the total, orbit-averaged change in $a$ is composed of 
two contributions (e.g., {\em Rubincam}, 1995, 1998; {\em Farinella et~al.},
1998; {\em Vokrouhlick\'y}, 1998a, 1999):
\begin{equation}
 \left(\frac{da}{dt}\right)_{\rm diurnal} = -\frac{8}{9} \frac{\alpha\Phi}{n}\, 
  W\left(R_\omega,\Theta_\omega\right)\,\cos\gamma\; , \label{yark_ana1}
\end{equation}
the diurnal effect, and
\begin{equation}
 \left(\frac{da}{dt}\right)_{\rm seasonal} =  \phantom{-}\frac{4}{9} \frac{\alpha\Phi}{n}\,
  W\left(R_n,\Theta_n\right)\,\sin^2\gamma\; , \label{yark_ana2}
\end{equation}
the seasonal effect. Here, $\Phi=\pi R^2 F/(mc)$, where $R$ is the radius of the body,
$F$ the solar radiation flux at the orbital distance $a$ from the Sun, $m$ is the mass of 
the body, $c$ is the light velocity, $n$ is the orbital mean motion and $\alpha=1-A$, 
with $A$ denoting the Bond albedo (e.g.,
{\em Vokrouhlick\'y and Bottke}, 2001). The $\Phi$ factor is characteristic to any
physical effect related to sunlight absorbed or scattered by the surface of the body.
Since $m\propto R^3$, one obtains a typical scaling $\Phi\propto 1/R$.

More importantly, the diurnal and seasonal components of
the Yarkovsky effect have a different dependence on the spin axis obliquity $\gamma$:
(i) the diurnal part is $\propto \cos\gamma$, and consequently can make a positive or negative
change of the semimajor axis, being maximum at $0^\circ$ and $180^\circ$ obliquity values, and 
(ii) the seasonal part is $\propto \sin^2\gamma$, and consequently always results in a 
decrease in semimajor axis, being maximum at $90^\circ$ obliquity. Their magnitude is
proportional to the function
\begin{equation}
 W\left(R_\nu,\Theta_\nu\right)= -\frac{\kappa_1(R_\nu)\,\Theta_\nu}{1+
  2\kappa_2(R_\nu)\,\Theta_\nu+\kappa_3(R_\nu)\,\Theta_\nu^2} \; , \label{yark_ana3}
\end{equation}
determined by the thermal parameters of the body and a frequency $\nu$. The latter is equal
either to the rotation frequency $\omega$ for the diurnal component, or the orbital mean
motion $n$ for the seasonal component. The thermal parameters required by the model
are: (i) the surface thermal conductivity $K$, (ii) the surface heat capacity $C$, and
(iii) the surface density $\rho$. These parameters, together with the frequency $\nu$, 
do not appear in (\ref{yark_ana3}) individually. Rather in the process of solving the heat
diffusion problem and determination of the orbital perturbations, 
they combine in two relevant parameters. First, they provide a scale length 
$\ell_\nu=\sqrt{K/(\rho C \nu)}$ which indicates a characteristic penetration depth
of temperature changes assuming the surface irradiation is periodic with the frequency $\nu$.
The non-dimensional radius of the body $R_\nu$ in Eq.~(\ref{yark_ana3}) is defined
by $R_\nu=R/\ell_\nu$. Secondly, the surface thermal inertia $\Gamma=\sqrt{K\rho C}$
enters the non-dimensional thermal parameter $\Theta_\nu$ in Eq.~(\ref{yark_ana3}) 
using a definition $\Theta_\nu=\Gamma\sqrt{\nu}/ (\epsilon \sigma T_\star^3)$,
with $\epsilon$ the thermal emissivity of the surface, $\sigma$ the Stefan-Boltzmann 
constant and $T_\star$ the sub-solar temperature ($\epsilon \sigma T_\star^4=\alpha
F$). When the characteristic size $R$ of the body is much larger than $\ell_\nu$ 
(a large-body limit), a situation met in the typical applications so far, the three 
$\kappa$-coefficients in Eq.~(\ref{yark_ana3}) are simply equal to $\case12$
({\em Rubincam}, 1995; see {\em Vokrouhlick\'y} (1998a)
for their behavior for an arbitrary value of $R_\nu$). Hence, for large bodies the
$W$-factors do not depend on the size $R$ and read $W\simeq W(\Theta_\nu)=-0.5\,\Theta_\nu/(1+
\Theta_\nu+0.5\,\Theta_\nu^2)$. Consequently, the Yarkovsky effect is maximum when $\Theta_\nu
\simeq 1$; for small or large values of $\Theta_\nu$ the effect vanishes. In this case,
the semimajor axis secular change $da/dt$ due to the Yarkovsky effect scales as 
$\propto 1/R$ with the characteristic radius $R$. For small asteroids, either
in the near-Earth space or in the main belt, $\Theta_\omega$ is typically of the order 
of unity (see also {\em Delb\`o et~al.}, this volume), while $\Theta_n$ is much 
smaller, which implies that the diurnal Yarkovsky component usually dominates
the seasonal component.

A handful of models were subsequently developed to probe the role of each of the
simplifying assumptions mentioned above using analytical, semi-analytical or fully
numerical methods. These include (i) an inhomogeneity of the thermal parameters (e.g.,
{\em Vokrouhlick\'y and Bro\v{z}}, 1999), (ii) a coupling of the diurnal and seasonal 
components of the Yarkovsky effect (e.g., {\em Vokrouhlick\'y}, 1999; {\em Sekiya and 
Shimoda}, 2013, 2014), (iii) effects of a non-spherical shape for simple (e.g., 
{\em Vokrouhlick\'y}, 1998b) or general geometries (including non-convex shapes and
the role of small-scale surface features; Sec.~2.2), (iv) a non-linearity of the 
surface boundary condition of the thermal model (e.g., {\em Sekiya and Shimoda}, 
2013, 2014), (v) the  role of very high orbital eccentricity (e.g., {\em Spitale 
and Greenberg}, 2001, 2002; {\em Sekiya and Shimoda}, 2014);
(vi) a non-principal axis rotation state (e.g., {\em Vokrouhlick\'y et~al.}, 2005a), or
(vii) the Yarkovsky effect for binary asteroids (e.g., {\em Vokrouhlick\'y et~al.}, 2005b).
Each of them was found to modify results from the zero-approximation model by as 
much as several tens of percent without modifying the fundamental dependence of 
the Yarkovsky effect on obliquity, size or thermal parameters (except perhaps for the 
special case of very high eccentricity orbits, where the sign of the Yarkovsky effect
may be changed; {\em Spitale and Greenberg}, 2001).
\smallskip

\noindent{\bf The YORP effect.-- }The YORP effect, the rotational counterpart of 
the Yarkovsky effect, broadly denotes the
torque arising from interaction with the impinging solar radiation. As in the orbital 
effect, the absorbed sunlight does not result in secular effects (e.g. {\em Breiter
et~al.}, 2007; {\em Nesvorn\'y and Vokrouhlick\'y}, 2008b; {\em Rubincam and 
Paddack}, 2010). Both directly scattered sunlight in the optical band and the 
recoil due to thermally reprocessed radiation, however, produce dynamical effects 
that accumulate over long timescales. In principle,
one would need to treat the two components of the YORP effect independently, since
the bi-directional characteristics of the scattered and thermally emitted radiation
are not the same and would produce different torques. Additionally, the thermal
component has a time lag due to the finite value of the surface thermal inertia and
its bi-directional function should formally depend of the time history of the particular
surface element. 

While these issues are at the forefront of current research (Sec.~2.2),
we start with a zero order approximation initially introduced by {\em Rubincam} (2000):
(i) the surface thermal inertia is neglected, such that thermal radiation
is re-emitted with no time lag, and (ii) the reflected and thermally radiated components
are simply assumed Lambertian (isotropic). This approximation avoids precise thermal 
modeling and the results are relatively insensitive to the body's surface albedo value. 
At face value, this looks simple, but 
layers of complexity unfold with the geometrical description of the surface. This is 
because the YORP effect vanishes for simple shape models (such as ellipsoids of 
rotation, {\em Breiter et~al.}, 2007) and stems from the irregular shape of the 
body (see already {\em Paddack}, 1969). Obviously, its quantitative 
description involves a near infinity of degrees of freedom if middle- to 
small-scale irregularities are included. This may actually be the case for real
asteroids because these irregularities may present a large collective cross-section 
and thus could dominate the overall strength of the YORP effect. This is now 
recognized as a major obstacle to our ability to model the YORP effect (Sec.~2.2).

The importance of fine details of geometry, somewhat unnoticed earlier, were unraveled by
the first analytical
and semi-analytical models of the YORP effect. There were two approaches developed
in parallel. {\em Scheeres} (2007) and {\em Scheeres and Mirrahimi} (2008) used the
polyhedral shape description as a starting point of their study, while {\em Nesvorn\'y 
and Vokrouhlick\'y} (2007, 2008a) and {\em Breiter and Micha{\l}ska} (2008) used 
shape modeling described by a series expansion in spherical harmonics. To keep
things simple, these initial models assumed principal axis rotation and disregarded
mutual shadowing of the surface facets. Both models predicted, after averaging the results
over the rotation and revolution cycles, a long-term change of the 
rotational rate $\omega$ and obliquity $\gamma$ (the precession rate effect is 
usually much smaller than the corresponding gravitational effect due to the Sun), 
which could be expressed as
\begin{equation}
 \frac{d\omega}{dt} = \frac{\Lambda}{C} \sum_{n\geq 1} A_n\,P_{2n}
  \left(\cos\gamma\right)\; , \label{yorp_ana1}
\end{equation}
and
\begin{equation}
 \frac{d\gamma}{dt} = \frac{\Lambda}{C\omega} \sum_{n\geq 1} B_n\,P^1_{2n}
  \left(\cos\gamma\right)\; . \label{yorp_ana2}
\end{equation}
Here, $\Lambda=2\,F R^3/(3c)$ with $C$ being the moment of inertia corresponding to the 
rotation axis (shortest axis of the inertia tensor), $P_{2n}(\cos\gamma)$ are the
Legendre polynomials of even degrees, and $P^1_{2n}(\cos\gamma)$ are the corresponding
associated Legendre functions. The particular characteristics of the even-degree
Legendre polynomials and Legendre functions of order 1 in Eqs.~(\ref{yorp_ana1}) and 
(\ref{yorp_ana2}) under prograde to retrograde reflection $\gamma \leftrightarrow \pi
- \gamma$ indicate the behavior of $d\omega/dt$ and $d\gamma/dt$: (i) the rotation-rate
change is symmetric, while (ii) the obliquity change is antisymmetric under this
transformation. Earlier numerical studies (e.g., {\em Rubincam}, 2000; {\em Vokrouhlick\'y
and \v{C}apek}, 2002; {\em \v{C}apek and Vokrouhlick\'y}, 2004) had suggested that
the net effect of YORP on rotation-rate often vanishes near $\gamma\sim 55^\circ$ and
$\gamma\sim 125^\circ$. This feature was finally understood using Eq.~(\ref{yorp_ana1})
because these obliquity values correspond to the roots of the second-degree Legendre
polynomial. The previous works that numerically treated smoothed surfaces thus mostly
described situations when the first term in the series played a dominant role.
When the effects of the surface finite thermal inertia are heuristically added to these
models, one finds that only the coefficients $B_n$ change (e.g., {\em Nesvorn\'y 
and Vokrouhlick\'y}, 2007, 2008a; {\em Breiter and Micha{\l}ska}, 2008). This confirms
an earlier numerical evidence of {\em \v{C}apek and Vokrouhlick\'y} (2004).

Since $C\propto R^5$, 
Eqs.~(\ref{yorp_ana1}) and (\ref{yorp_ana2}) imply that both rotation-rate and 
obliquity effects scale with the characteristic radius as $\propto 1/R^2$. This is an 
important difference with respect to the ``more shallow'' size dependence of the Yarkovsky 
effect, and it implies that YORP's ability to change the rotation state
increases very rapidly moving to smaller objects. Additionally, we understand well that
for very small bodies the Yarkovsky effect becomes eventually nil. When
the characteristic radius $R$ becomes comparable to the penetration depth $\ell_\omega$
of the diurnal thermal wave the efficient heat conduction across the volume of the
body makes temperature differences on the surface very small. However, {\em Breiter et~al.}
(2010a) suggested that in the same limit the YORP strength becomes $\propto 1/R$, still
increasing for small objects. Additionally, their result was only concerned with the
thermal component of the YORP effect, while the part related to the direct sunlight
scattering in optical waveband continues to scale with $\propto 1/R^2$. Thus, 
the fate of small meteoroids' rotation is still unknown at present.

The principal difference in complexity of the YORP effect results in Eqs.~(\ref{yorp_ana1}) 
and (\ref{yorp_ana2}), as compared to simple estimates in Eqs.~(\ref{yark_ana1}) 
and (\ref{yark_ana2}) for the Yarkovsky effect, is their infinite series nature. 
The non-dimensional coefficients $A_n$ and $B_n$ in Eqs.~(\ref{yorp_ana1}) and 
(\ref{yorp_ana2}) are determined by the shape of the body, either
analytically or semi-analytically (e.g., {\em Nesvorn\'y and Vokrouhlick\'y}, 
2007, 2008a; {\em Scheeres and Mirrahimi}, 2008; {\em Breiter and Micha{\l}ska},
2008; {\em Kaasalainen and Nortunen}, 2013). 
Interestingly, analytical methods help to understand that torque component that 
changes the spin rate and the components that change the axis orientation couple, at 
leading order, to different attributes of the surface.
The spin torque couples to chirality -- the difference between eastward and westward facing
slopes -- while the other components couple merely to asphericity. Mathematically, 
this concerns the symmetric and antisymmetric terms in the Fourier expansion of the 
topography. If mutual shadowing of the
surface facets is to be taken into account, one may use a semi-analytic
approach mentioned by {\em Breiter et~al.} (2011); see already {\em Scheeres and
Mirrahimi} (2008). Depending on details of the shape, the series in 
(\ref{yorp_ana1}) and (\ref{yorp_ana2}) may either converge quickly, with the 
first few terms dominating the overall behavior, or may  slowly converge, with 
high-degree terms continuing to contribute (e.g., {\em Nesvorn\'y and 
Vokrouhlick\'y}, 2007, 2008a; {\em Kaasalainen and Nortunen}, 2013). 

While this behavior had been noticed in analytical modeling, a detailed numerical
study of YORP sensitivity on astronomically-motivated, small-scale surface
features such as craters and/or boulders was performed by {\em Statler} (2009).
This also allowed {\em Statler} (2009) to suggest a new direction to 
YORP studies. He noted that the sensitivity of YORP on such small scale features 
may affect its variability on short-enough timescales to significantly modify 
the long-term evolution of rotation rate, with the evolution changing from a 
smooth flow toward asymptotic state to a random walk (Sec. 2.2). 

The quadrupole ($2n=2$), being the highest multipole participating in the
series expansion in Eqs.~(\ref{yorp_ana1}) and (\ref{yorp_ana2}), is related
to the assumption of coincidence between the reference frame origin and
the geometric center of the body (i.e., its center-of-mass for homogeneous 
density distribution). If instead the rotation
axis is displaced from this point, additional terms in the series become
activated and the coefficients $(A_n,B_n)$ become modified, and thus the predicted
YORP torque will change (e.g., {\em Nesvorn\'y and Vokrouhlick\'y}, 2007, 2008a).
This theoretical possibility has found an interesting geophysics interpretation
for (25143) Itokawa's anomalously small YORP value by {\em Scheeres and Gaskell}
(2008) (see Sec.~3.2, {\em Breiter et~al.}, 2009, and eventually {\em 
Lowry et~al.}, 2014).

\bigskip
\noindent
 \textbf{2.2 Frontiers in modeling efforts}
\bigskip

\noindent{\bf Resolved and unresolved surface irregularities.-- }While 
the models discussed above suffice to 
describe broad-scale features of the Yarkovsky and YORP effects, there 
are important aspects which are intrinsically 
{\em nonlinear}. Current models need to explicitly treat these nonlinearities
in order to capture the physical essence of radiation recoil mechanisms and
to provide precise predictions. Here we discuss some recent efforts along
these lines.

The simplest of such nonlinear effects is {\em shadowing\/} of some parts of the
surface by other parts, which can occur on surfaces that are not
convex. By blocking the Sun, shadowing lowers the incident flux, and
increases the temperature contrast, compared to the clear-horizon
case. Computationally, shadowing requires testing whether the
sunward-pointing ray from each surface element intersects another surface
element (e.g., {\em Vokrouhlick\'y and \v{C}apek}, 2002). This ``who blocks 
whom'' problem is of ${\cal O}(N^2)$ complexity (where $N$ is the number of
surface elements); but there are strategies for storing an initial ${\cal O}
(N^2)$ calculation so that all subsequent calculations are only ${\cal 
O}(N)$ (e.g., {\em Statler}, 2009).

Closely related to shadowing are the processes of {\em self-heating}
(e.g., {\em Rozitis and Green}, 2013); these can be
split conceptually into {\em self-illumination}, in which a surface element
absorbs reflected solar flux from other parts of the surface, and
{\em self-irradiation}, where it absorbs re-radiated thermal infrared.
Self-heating has the tendency to reduce the temperature contrast, by
illuminating regions in shadow. Computing these effects requires prescriptions
for the angular distribution of reflected and re-radiated power from
an arbitrary surface element, as well as the solution to the ``who
sees whom'' problem -- similar to the ``who blocks whom'' problem from
shadowing. But since energy is traded between pairs of surface elements,
self-heating, unlike shadowing, is unavoidably ${\cal O}(N^2)$ if full 
accuracy is required.

As mentioned in Sec.~2.1, a periodic driving at a frequency $\nu$ 
introduces a length scale, the thermal skin depth $\ell_\nu$. Asteroid 
surfaces are driven quasi-periodically, with the fundamental modes at
the diurnal and seasonal frequencies. For typical materials, $\ell_\nu$ 
is of the order of meters for the seasonal cycle and millimeters to centimeters 
for the diurnal cycle. If the surface's radius of curvature $s$ satisfies 
the condition $s \gg \ell_n$, one can consider surface elements to be 
independent (facilitating parallelization) and solve the heat conduction
problem as a function of the depth only. The radiated flux then depends on 
the material parameters only through the thermal inertia $\Gamma$. Most 
models that treat conduction explicitly do so in such 1D approximation. Standard
finite-difference methods are typically used to find a solution over
a rotation or around a full orbit; but numerical convergence
can be slow (though acceleration schemes were also considered, {\em
Breiter et~al.}, 2010b). Whether the condition $s \gg \ell_n$ is truly satisfied
depends on the scale on which topography is resolved. A surface boulder
can give an object a locally small radius of curvature and 3D effects may
become important. Full 3D conduction is computationally expensive (e.g., 
{\em Golubov et~al.}, 2014; {\em \v{S}eve\v{c}ek et~al.}, 2015), but the 
potential consequences are significant. In this case, a general finite-element
method is used to solve the heat diffusion problem.

{\em Surface roughness} concerns the effects of unresolved texture on
reflection, absorption, and re-radiation. Parametric models for a rough-surface
reflectance are well developed (e.g., {\em Hapke}, 1993, and references
therein; {\em Breiter and Vokrouhlick\'y}, 2011, in an application to the YORP
effect), though the functional forms and parameter values are matters 
of current research. 
Models for the thermal emission are at present purely numerical. In the most 
complete implementation ({\em Rozitis and Green}, 2012, 2013), a 
high-resolution model of a crater field 
is embedded inside a coarse-resolution model of a full object. The primary 
effects of roughness in this model are to enhance the directionality 
(``beaming'') of the radiated intensity (relative to Lambertian emission), 
and to direct the radiated momentum slightly away from the surface normal, 
toward the Sun. Roughness models for emission and for reflection are not 
automatically mutually consistent; and the emission models employ the 1D 
approximation for heat conduction despite the likelihood that $s$ may not 
be much larger than $\ell_\omega$ at the roughness scale.

Finally, nonlinear {\em dynamical coupling\/} affects both spin
evolution and the orbital drift modulated by the spin state. Yarkovsky
evolution models have generally incorporated heuristic prescriptions based
on the YORP cycle (e.g., {\em Rubincam}, 2000; {\em Vokrouhlick\'y and
\v{C}apek}, 2002), with possibly important effects of spin-induced material 
motion or reshaping included only in rudimentary ways. These processes 
may be modeled with particle-based discrete-element numerical codes 
(e.g., {\em Richardson et~al.}, 2005; {\em Schwartz et~al.}, 2012) and
semi-numerical granular dynamics in pre-defined potential fields
(e.g., {\em Scheeres}, 2015). Simulated rubble piles artificially fed 
with angular momentum are seen to reshape and shed mass (e.g., {\em
Walsh et~al.}, 2008; {\em Scheeres}, 2015). Linking a particle code
with a thermophysical YORP model would then allow the coupled spin and
shape evolution to be followed self-consistently.

{\em Statler} (2009) argued that topographic sensitivity would make rubble
piles, or any objects with loose regolith, susceptible to possibly large changes
in torque triggered by small, centrifugally driven changes in shape.
Repeated interruptions of the YORP cycle might then render the overall
spin evolution stochastic and significantly extend the timescale of the YORP
cycles ({\em self-limitation\/} property of YORP). {\em Cotto-Figueroa et~al.}
(2015) have tested this prediction by simulating self-consistently the coupled
spin and shape evolution (togging between configurations in a limit cycle),
and stagnating behaviors that result in YORP self-limitation. {\em Bottke et~al.}
(2015) implemented a heuristic form of such stochastic
YORP in a Yarkovsky drift model to find an agreement with the
structure of the Eulalia asteroid family.

Accurate Yarkovsky measurements allow constraining mass and bulk
density (Sec.~4.1), but rely on precise models, with an important
component due to the surface features discussed above. {\em Rozitis 
and Green} (2012) show that surface roughness can increase the Yarkovsky 
force by tens of percent, owing mainly to the beaming. Including the seasonal 
effect caused by the deeper-penetrating thermal wave can have a comparable 
influence. Self-heating, in contrast, has a minimal influence on Yarkovsky 
forces (e.g., {\em Rozitis and Green}, 2013). On the other hand, the same
works indicate that the YORP effect is in general dampened by beaming
because it equalizes torques on opposite sides of the body.

{\em Golubov and Krugly} (2012) highlight another small-scale aspect of 
the YORP effect: an asymmetric heat conduction across surface features for which 
$s \lesssim \ell_\omega$. A rock conducts heat from its
sunlit east side to its shadowed west side in the morning, and from its
west side back to its east side in the afternoon. Owing to nighttime
cooling, the morning temperature gradient is steeper, and hence more heat
is conducted to, and radiated from, the west side, resulting in an eastward
recoil. Clearly, if the collective cross-section of such surface features
is large, details of conduction across them may have significant
consequences.
Ideally, the situation calls for a complete 3D heat transfer model
(e.g., {\em Golubov et~al.}, 2014; {\em \v{S}eve\v{c}ek et~al.}, 2015). 
Importantly, these studies indicate an overall tendency for YORP to spin
objects up. However, a better understanding of small-scale surface 
effects is essential to understand YORP's long-term dynamics.
\smallskip

\noindent{\bf Time domain issues (tumbling).-- }A particular problem 
in the modeling of the thermal effects occurs for
tumbling bodies. This is because solving the heat diffusion in the
body involves also the time domain. While the spatial dimensions
are naturally bound, the time coordinate is not in general. 
However, both analytical and numerical methods involve
finite time domains: the analytical approaches use a development in
the Fourier series, while the effective numerical methods use iterations
that require one to identify configurations at some moments in time.
For bodies rotating about the principal axis of the inertia
tensor, thus having a fixed direction in the inertial space, it is
usually easy to modify the rotation period within its uncertainty
limits such that it represents an integer fraction of the orbital
period. The orbital period is then the fundamental time interval
for the solution. This picture becomes more complicated for
tumbling objects whose rotation is not characterized by a single
time period. Rather, it is fully described with two periods,
the proper rotation period and precession period, which may not
be commensurable.

This situation has been numerically studied by {\em Vokrouhlick\'y
et~al.} (2005a) in the case of (4179)~Toutatis, and more recently
also in the case of (99942)~Apophis by {\em Vokrouhlick\'y et~al.} (2015). 
Both studies suggest the tumbling may not necessarily ``shut down the 
Yarkovsky effect'', at least in the large-bodies regime. Rather, 
it has been found that the Yarkovsky acceleration for these tumbling 
objects is well represented by a simple estimate valid for bodies 
rotating about the shortest axis of the inertia tensor in a direction 
of the rotational angular momentum and with the fundamental period 
of tumbling, generally the precession period. 
\smallskip

\noindent{\bf More than one body (binarity).-- }Another particular 
case is the Yarkovsky effect for binaries (see also {\em Vokrouhlick\'y
et~al.}, 2005b). Unless the satellite has nearly
the same size as the primary component, the rule of thumb is that
the heliocentric motion of the system's center-of-mass is affected primarily
by the Yarkovsky acceleration of the primary component. The motion of the
satellite feels rather the Yarkovsky acceleration of the satellite itself.
Nevertheless, a secular change in the orbit of the satellite is actually caused
by an interplay of the thermal effects and the shadow geometry in
the system dubbed the Yarkovsky-Schach effect (and introduced years ago in
space geodesy; {\em Rubincam}, 1982). However, it turns out that the BYORP 
effect, discussed in Sec.~2.3, is more important and dominates the orbital
evolution of the satellite.

\bigskip
\noindent
 \textbf{2.3 Binary YORP}
\bigskip

The Binary YORP (BYORP) effect was first proposed in a paper by 
{\em \'Cuk and Burns} (2005). They noted that an asymmetrically-shaped 
synchronous secondary asteroid in a binary system should be subject to 
a net force differential that acts on average in a direction tangent 
to the orbit. Thus, as the secondary orbits about the primary body
and maintains synchronicity, 
this would lead to either an acceleration or deceleration of the 
secondary which would cause the mutual orbit of the system to spiral 
out or in, respectively. This seminal paper presented a basic conceptual 
model for the BYORP effect and provided a broad survey of many of the 
possible implications and observable outcomes of this effect. It also 
numerically studied the evolution of randomly shaped secondary bodies 
over a year to establish the physical validity of their model. 
It is key to note that a necessary condition for the BYORP effect is 
that at least one of the bodies be synchronous with the orbit, and 
it can be shut off if both bodies are non-synchronous. {\em \'Cuk and 
Burns} concluded that the BYORP effect should be quite strong and lead 
binary asteroids to either spiral in towards each other or cause them to 
escape in relatively short periods of time. This was further expanded in 
a second paper by {\em \'Cuk} (2007) that outlined significant implications 
for the rate of creation and destruction of binary asteroid systems in 
both the NEA and main belt population, leading to the initial estimate 
of binary asteroid lifetimes due to BYORP on the order of only $100$~ky. 

{\em McMahon and Scheeres} (2010a,b) then developed a detailed 
analytical model of the BYORP effect that utilized the existing shape model 
of the (66391) 1999~KW4 binary asteroid satellite ({\em Ostro et~al.}, 2006).
In their approach the solar radiation force was mapped into the secondary-fixed 
frame and expanded as a Fourier series, following a similar approach to the 
YORP model development of {\em Scheeres} (2007). This enables any given 
shape model to be expressed with a series of coefficients that can be 
directly computed, and allows for time averaging. Using this approach they 
showed that the 
primary outcome of the BYORP effect could be reduced to a single parameter 
--the so-called ``BYORP coefficient'' $B$-- uniquely computed from a 
given shape model. Henceforth, if the secondary is in a near-circular orbit,
the entire BYORP effect results in simple evolutionary equations for semi-major
axis $a$ and eccentricity $e(\ll 1)$ of the binary orbit
\begin{eqnarray}
 \frac{da}{dt} & = & \phantom{-}\frac{F B}{c \eta^\prime}
   \frac{a^{3/2}}{m_2 \sqrt{\mu}} \; , \label{eq:BYORP_a} \\
 \frac{de}{dt} & = & - \frac{F B}{4c\eta^\prime}
  \frac{e\, a^{3/2} }{m_2 \sqrt{\mu}} \; , \label{eq:BYORP_b}
\end{eqnarray}
where again $F$ is the solar radiation flux at the heliocentric distance
$a^\prime$ (equal to the semimajor axis of the heliocentric orbit), $\eta^\prime
=\sqrt{1-e^{\prime\,2}}$ with $e^\prime$ being the eccentricity of the 
heliocentric orbit, $c$ is the light velocity,
$m_2$ is the mass of the secondary, and $\mu = G\,(m_1+m_2)$ is the gravitational 
parameter of the binary system. If the orbit is expansive ($B  > 0$), the eccentricity 
will be stabilized, and vice-versa (see already {\em \'Cuk and Burns}, 2005). 
In the case where the binary orbit is highly elliptic, the evolutionary equations become 
much more complex, and require additional Fourier coefficients to be included into 
the secular equations, as discussed in detail in {\em McMahon and Scheeres} (2010a). 

\begin{figure}[t]
 \epsscale{1.}
 \plotone{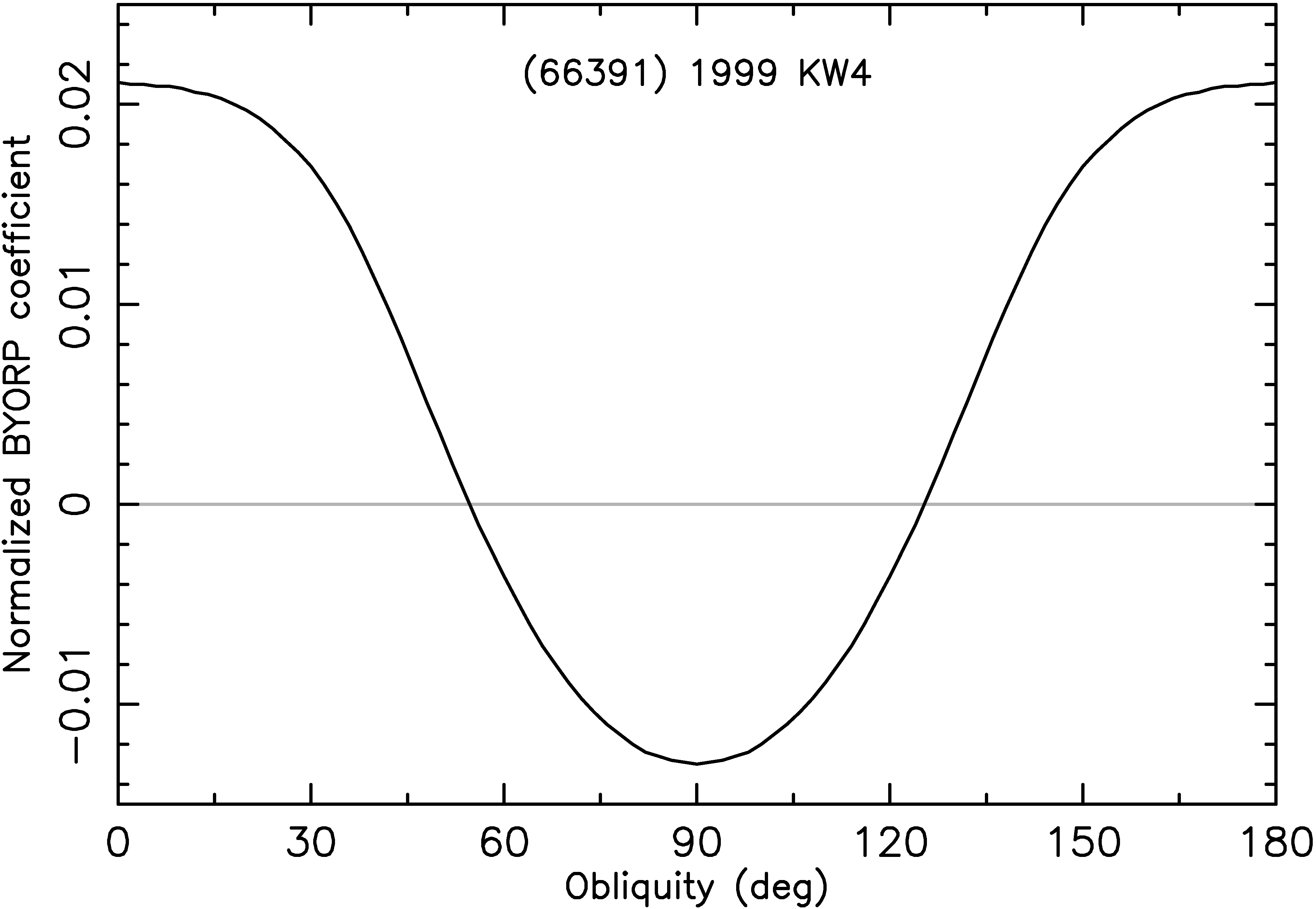}
 \caption{\small BYORP coefficient $B$, normalized by the square of the
 effective radius, computed for the secondary of the (66391) 1999~KW4 binary
 asteroid system, as a function of the binary orbital obliquity (abscissa).}
 \label{fig_byorp}
\end{figure}

The BYORP coefficient $B$ is computed as a function of the shape of the body and 
the obliquity of the binary's orbit relative to the heliocentric orbit of 
the system. Assume a model for the instantaneous solar radiation force acting 
on the secondary has been formulated by some means, denoted as ${\bf F}_{SRP}(M,M^\prime)$, 
where $M$ and $M^\prime$ are the mean anomalies of the binary mutual orbit and 
heliocentric orbit, respectively. Then the computation of the BYORP coefficient
requires double averaging of the radiation force over the binary and
heliocentric revolution cycles, and projection in the
direction of binary orbital motion (denoted here in abstract as $\hat{\bf t}$):
\begin{equation}
 B  =  \hat{\bf t} \cdot \frac{1}{(2\pi)^2} \int_{0}^{2\pi} \int_{0}^{2\pi}
  \frac{{\bf F}_{SRP}}{P(r_s)} \ dM \ dM^\prime \; ,
\end{equation}
where $P(r_s)=(F/c)\,(a^\prime/r_s)^2$ is the solar radiation pressure acting on 
the unit surface area of the body at the heliocentric distance $r_s$. The 
normalization
by $P$ implies that units of the BYORP coefficient are measured in area; thus
$B$ can be further normalized by dividing it by the effective radius squared 
of the secondary body. The BYORP coefficient is a function of several physical 
quantities such as albedo, surface topography, and potentially thermo-physical 
effects. However, the strongest variation of the BYORP coefficient is seen to vary 
with the binary obliquity with respect to the heliocentric orbit (Fig.~\ref{fig_byorp}). 
If the synchronous body is rotated by 180$^\circ$ relative to the orbit, then the 
sign of the BYORP coefficient will be uniformly reversed. Due to this, when a 
body initially enters into a synchronous state it is supposed that the 
probability of it being either positive or negative is 50\%. 

A more recent analysis of the BYORP effect was published by {\em Steinberg and Sari}
(2011) who found a 
positive correlation between the strength of the BYORP and YORP effects for 
bodies, and provided predictions related to the BYORP-driven evolution of the 
obliquity of a binary asteroid. In addition, they probed the possible effects of 
thermophysical models on the evolution of a binary system.

The above discussions focus on the effect of BYORP in isolation, and not in 
conjunction with other evolutionary effects. Recent work though has found 
that the BYORP effect can mix with other evolutionary effects in surprising 
ways that require additional verification and study. These are primarily discussed 
later in Sec.~5.3, where the long-term evolution of binary systems subject to 
BYORP is briefly considered. However, one of these combined effects has 
significant implications and is discussed here.

In particular, {\em Jacobson and Scheeres} (2011b) proposed existence of an 
equilibrium between the BYORP effect and tides. For this equilibrium to exist the BYORP 
coefficient must be negative, leading to a contractive system, and the primary 
asteroid must be spinning faster than the orbit rate. This creates a tidal 
dissipation torque that acts to expand the secondary orbit. Based on current 
theories of energy dissipation within rubble pile asteroids (e.g., 
{\em Goldreich and Sari}, 2009),  {\em Jacobson and Scheeres} noted
that all singly-synchronous rubble pile binary asteroids with a negative 
BYORP coefficient for the secondary should approach a stable equilibrium 
that balances these two effects. This is significant as it provides  
mechanism for BYORP's persistent effect to become stalled, leaving binary 
asteroids that should remain stable over long time spans. This, in turn, means 
that rapid formation of binary asteroids is not needed to explain the 
current population.  
\bigskip

\centerline{\textbf{3. DIRECT DETECTIONS}}
\bigskip

Accurate observations have now allowed direct detections of both the
Yarkovsky and YORP effects. This is an important validation of their
underlying concepts, but also it motivates further
development of the theory. These direct detections have two aspects
of usefulness or application. First, the Yarkovsky effect is being
currently implemented as a routine part of the orbit determination
of small near-Earth asteroids whose orbits are accurately constrained
in the forefront software packages. Additionally, the Yarkovsky effect 
is already known to be an essential part of the Earth impact hazard
computations in selected cases (Sec.~4.2 and {\em Farnocchia et~al.},
this volume). Second, many applications of the Yarkovsky
and YORP effects involve statistical studies of small body populations
in the Solar system rather than a detailed description of the dynamics of 
individual objects. Aside from a general validation, the known detections
help in setting parameter intervals that could be used in these statistical
studies.
\begin{figure}[t]
 \epsscale{1.}
 \plotone{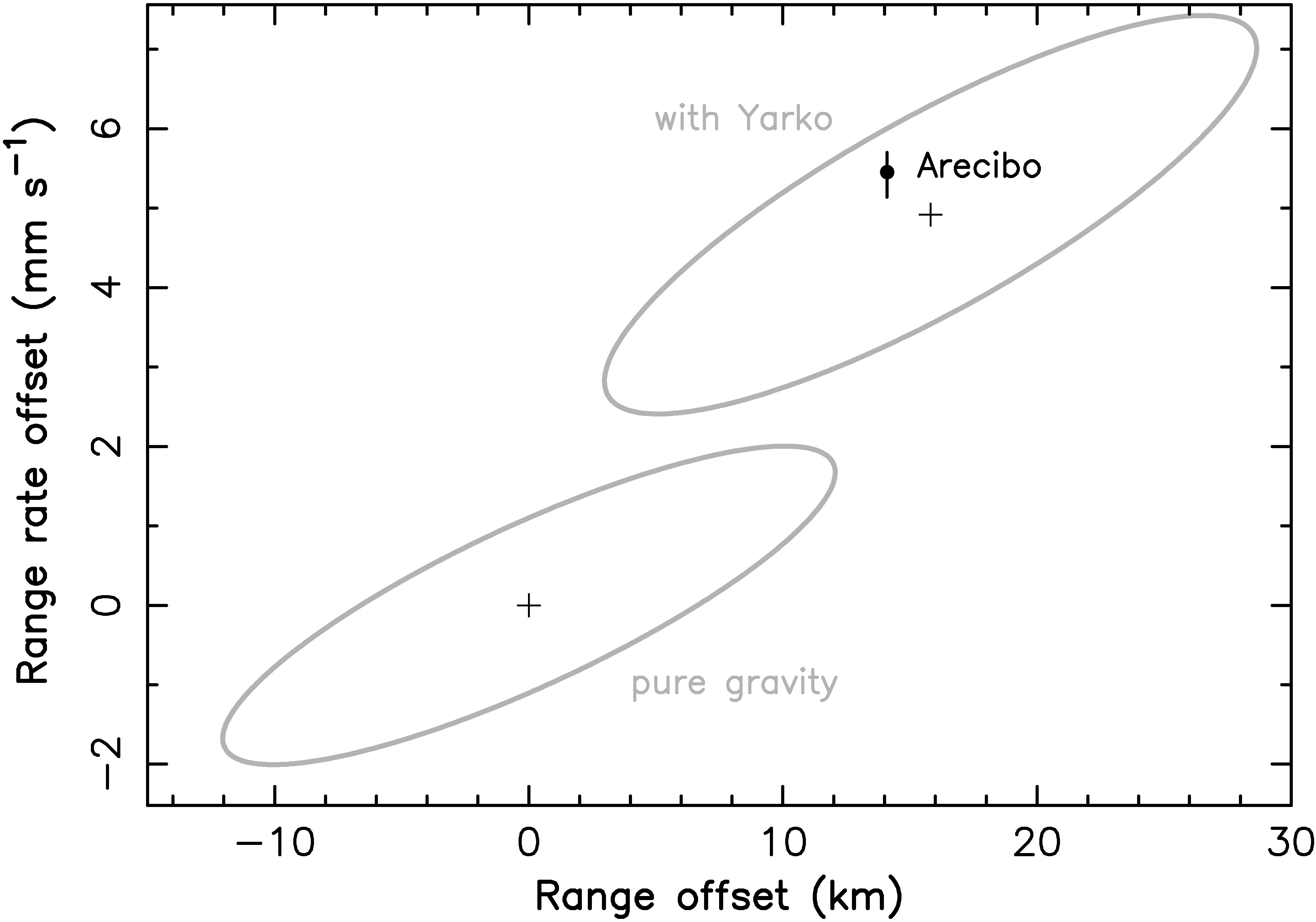}
 \caption{\small Orbital solution of near-Earth asteroid (6489)
 Golevka from astrometric data before May 2003 projected into the
 plane of radar  observables: (i) range at the abscissa, and (ii)
 range-rate on the ordinate. The origin referred to the center of
 the nominal solution that only includes gravitational perturbations.
 The gray ellipse labeled ``pure gravity'' represents a 90\%
 confidence level in the orbital solution due to uncertainties in astrometric
 observations as well as small body and planetary masses. The center
 of the gray ellipse labeled ``with Yarko'' is the predicted solution
 with the nominal Yarkovsky forces included (taken from {\em Vokrouhlick\'y
 et~al.}, 2000);
 note the range offset of $\sim 15$~km and the range rate offset of
 $\sim 5$~mm/s. The actual Arecibo observations from May 24, 26 and 27,
 2003 are shown by the black symbol (the measurement uncertainty in range
 is too small to be noted in this scale). The observations fall
 perfectly in the uncertainty region of the orbital solution
 containing the Yarkovsky forces. Adapted from {\em Chesley et~al.} (2003).}
 \label{fya}
\end{figure}
\begin{figure}[t]
 \epsscale{1.}
 \plotone{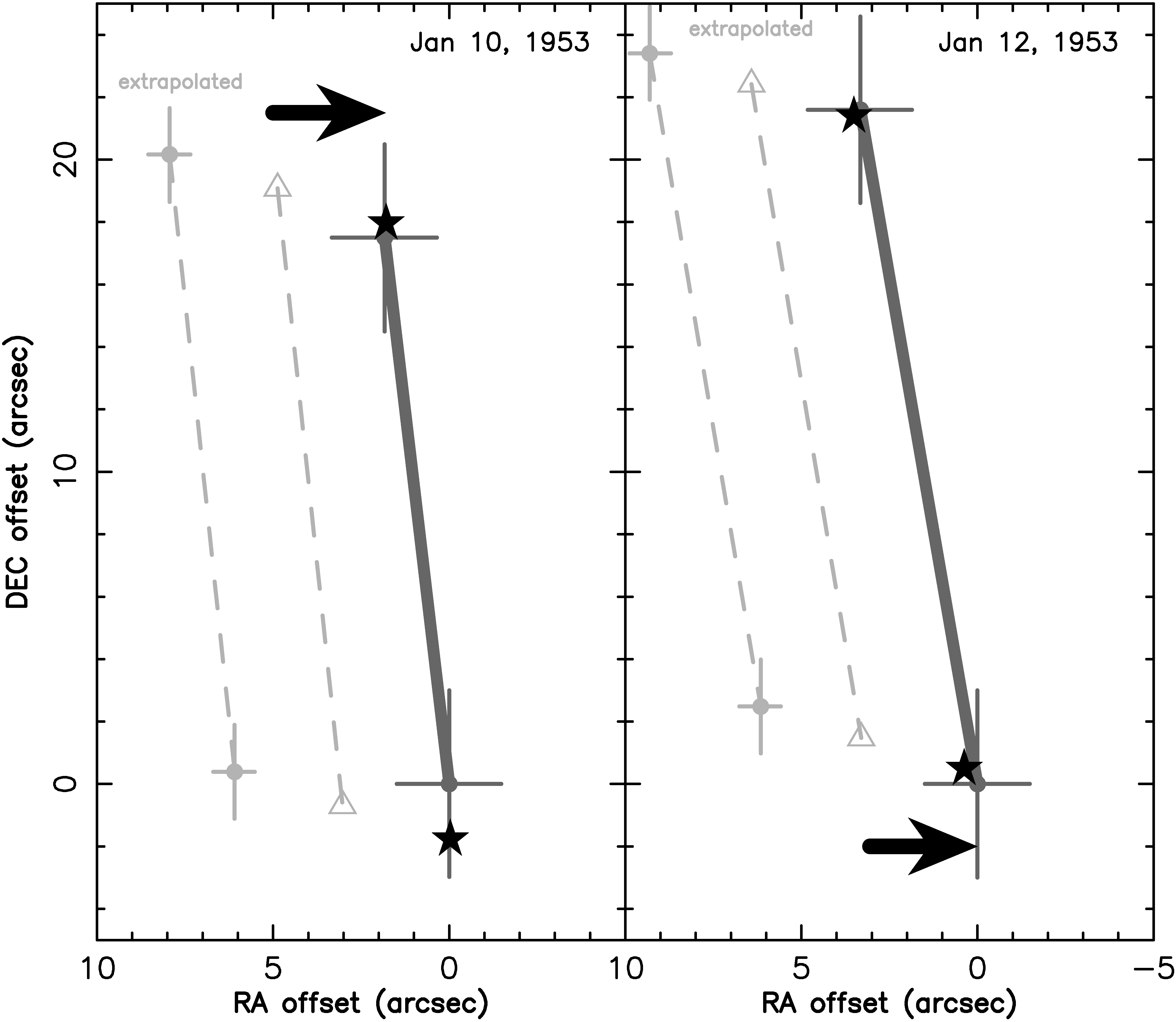}
 \caption{\small Measured and predicted positions of (152563) 1992 BF on
 January 10 (left) and 12 (right), 1953. The dark gray solid line is the 
 asteroid trail appearing on Palomar plates on the two nights. Coordinate
 origin, right ascension at the abscissa and declination at the ordinate,
 is arbitrarily set to the end of the respective trail. The leftmost dashed
 trail labeled ``extrapolated'' represent pure extrapolation of the
 modern orbit without the thermal forces included. The mismatch in
 right ascension slightly improves if the 1953 data are included in
 the orbital solution as shown by the middle dashed trail. Still, the solution
 is more than 3$\sigma$ away from the measured trail. Only when the thermal
 accelerations are included in the orbital solution, do the predicted orbital
 position matches the observations: stars show fitted position at the beginning
 and the end of the trail. Adapted from {\em Vokrouhlick\'y et~al.} (2008).}
 \label{fya_bf}
\end{figure}

\bigskip
\noindent
\textbf{3.1 Yarkovsky effect}
\bigskip

The possibility of detecting the Yarkovsky effect as a measurable orbital deviation
was first proposed by {\em Vokrouhlick\'y et~al.} (2000). The idea is at first 
astounding given that the transverse thermal recoil force on a half-kilometer 
near-Earth asteroid (NEA) should be at most $0.1$~N, causing an acceleration of only 
$\sim 1\; \mathrm{pm/s}^2$. And yet such small perturbations can lead to tens of 
kilometers of orbital deviation for half-kilometer NEAs after only a decade. In 
principle, such a deviation is readily detectable during an Earth close approach, 
either by optical or radar observations, but the key challenge is that the 
precision of the position prediction must be significantly smaller than the 
Yarkovsky deviation that is to be measured. In practical terms this means that 
detection of the Yarkovsky effect acting on a typical half-kilometer NEA 
requires at least three radar ranging apparitions spread over a decade, or 
several decades of optical astrometry in the absence of radar ranging. Of course 
smaller objects could in principle reveal the Yarkovsky effect much more quickly, 
but the problem for small objects is that it is more difficult to build up suitable 
astrometric datasets. Because of this, only a few objects with diameters $D< 100$~m 
have direct detections of the Yarkovsky effect.

It should be pointed out that observations do not allow to measure the secular 
change in the orbital semimajor axis directly. Rather, they reveal an associated 
displacement in the asteroid position along the orbit, an effect that progresses 
$\propto t^2$ in a given time $t$ (see {\em Vokrouhlick\'y et~al.}, 2000). This 
is similar to the way how the YORP effect is observed as discussed in Sec.~3.2.

As predicted by {\em Vokrouhlick\'y et~al.} (2000), (6489)~Golevka was the first 
asteroid with an unambiguous detection of the signature of the Yarkovsky effect 
in its orbit ({\em Chesley et~al.}, 2003). In this case the detection was possible 
only due to the availability of three well-separated radar ranging apparitions, in 
1991, 1995 and 2003. The first two radar apparitions constrain the semimajor axis, 
affording a precise position prediction in 2003, while the 2003 radar ranging 
revealed a deviation from a ballistic trajectory. Figure~\ref{fya} depicts the 
predicted 2003 delay-Doppler observations with their uncertainty along with the 
associated uncertainties. The predictions were well separated with $>90\%$ confidence, 
and the actual asteroid position fell close to the Yarkovsky prediction.

The second reported detection of the Yarkovsky effect was for (152563) 1992~BF, which 
was also the first detection that did not rely on radar astrometry ({\em Vokrouhlick\'y
et~al.}, 2008). This half-kilometer asteroid had a 13-year optical arc (1992-2005) and 
four archival positions over two nights dating to 1953. These so-called precovery 
observations could not be fit to a purely gravitational orbit, but including the 
Yarkovsky effect in the orbit fitting enabled the observations to fit well and allowed 
a $da/dt$ estimate with the signal-to-noise ratio SNR $\simeq 15$ (Fig.~\ref{fya_bf}). In 
these cases, where the detection relies heavily on isolated and archival data, caution 
is warranted to avoid the possibility that mis-measurement or astrometric time tag 
errors are corrupting the result. As depicted in Fig.~\ref{fya_bf}, the 1953 
position offsets could not be attributed to timing errors, and the trail positions 
were remeasured with modern catalogs.

In subsequent studies a progressively increasing number of Yarkovsky detections have 
been announced ({\em Chesley et~al.}, 2008; {\em Nugent et~al.}, 2012a; {\em Farnocchia 
et~al.}, 2013b). The most precise Yarkovsky measurement is that of (101955)~Bennu, 
the target of the OSIRIS-REx asteroid sample return mission, which has a $0.5$\% 
precision Yarkovsky detection, by far the finest precision reported to date. At the 
extremes, asteroid 2009~BD is the smallest object ($D\sim 4$~m) with a verified 
Yarkovsky detection, which was achieved because of its Earth-like orbit and the 
two-year arc of observations that the orbit enabled ({\em Mommert et~al.}, 2014). On 
the large end, there are two detections with diameter $2$ to $3$~km, namely 
(2100)~Ra-Shalom and (4179)~Toutatis ({\em Nugent et~al.}, 2012a; {\em Farnocchia 
et~al.}, 2013b), which are both exceptionally well observed, having 4 and 5 radar 
apparitions, respectively.

To initially test for a signal from the Yarkovsky effect in the astrometric data of 
a given object one can fit the orbit with a transverse nongravitational acceleration 
$a_T = A_2/r^2$, with $A_2$ being an estimated parameter, in addition to the orbital 
elements. This simple model yields a mean semimajor axis drift rate proportional 
to $A_2$, thus capturing the salient orbital deviation due to the Yarkovsky effect. 
The approach of using a one-parameter ($A_2$) Yarkovsky model is particularly 
convenient because it completely bypasses the thermophysical processes that are 
otherwise fundamental to the Yarkovsky effect. Instead, by focusing only on the level of 
perturbation visible in the orbit, one is able to discern the Yarkovsky effect 
in absence of any knowledge of physical properties. And yet, as we shall see in 
Sec.~4.1, the detection of a Yarkovsky drift can be used to estimate or infer 
a number of the physical and dynamical characteristics of the body. Obviously,
in the case of bodies with particular interest, one can use a detailed
thermophysical model of the Yarkovsky acceleration for the orbit determination
in a subsequent analysis.

A population-wise, head-on approach to Yarkovsky detection thus starts with 
the list of asteroids with relatively secure orbits, e.g., at least 
$100$~days of observational arc, among the NEAs. For 
each considered object the statistical significance of the Yarkovsky effect is 
obtained from the estimated value of $A_2$ and its {\it a posteriori} uncertainty 
$\sigma_{A_2}$ according to ${\rm SNR}= |A_2|/\sigma_{A_2}$, where ${\rm SNR}> 3$ is generally 
considered to be a significant detection. Another parameter that is helpful in 
interpreting the results for a given object is the ratio between the estimated 
value of $A_2$ and the expected value for extreme obliquity and the known or 
inferred asteroid size, which we call ${A_2}_{\rm max}$. The value of 
${A_2}_{\rm max}$ can be obtained by, for instance, a simple diameter scaling from 
the Bennu result ({\em Farnocchia et~al.}, 2013b; {\em Chesley et~al.}, 2014). The 
ratio ${\cal S} = A_2/
{A_2}_{\rm max} = \mathrm{SNR}/\mathrm{SNR}_{\rm max}$ provides an indication of 
how the estimated value of $A_2$ compares to what could be theoretically 
expected. A value of ${\cal S} \gg 1$ indicates that the transverse 
nongravitational acceleration may be too strong to be related to the Yarkovsky
effect. 
This could imply that the body has a far smaller density or size than assumed, 
or that nongravitational accelerations other than Yarkovsky are at play. A 
large value of $\cal S$ could also imply a spurious $A_2$ estimate due to 
corrupt astrometry in the orbital fit. On the other hand, ${\cal S} \ll 1$ 
would suggest the possibility of higher density, size or surface thermal inertia
than assumed, but is often more readily explained by mid-range obliquity, which 
tends to null the diurnal component of the Yarkovsky drift.
\begin{figure}[t]
 \epsscale{1.}
 \plotone{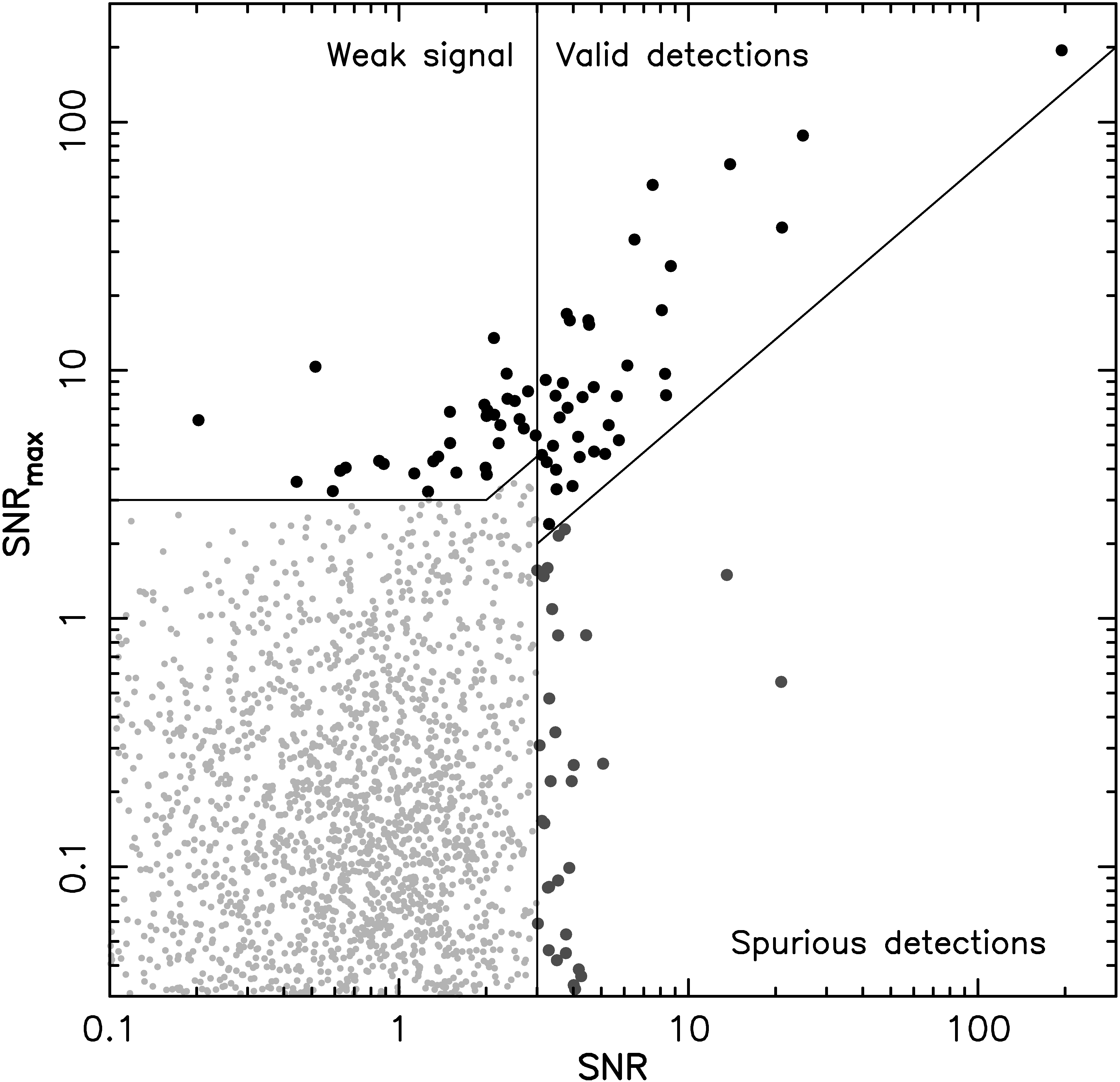}
 \caption{\small ${\rm SNR} = A_2/\sigma_{A_2}$,
  with $A_2$ being the parameter of an empirical transverse acceleration
  and $\sigma_{A_2}$ its formal uncertainty, for reliable orbits of
  NEAs at the abscissa. The ordinate shows ${\rm SNR}_{\rm max}$,
  the maximum expected value of SNR for the body (from an estimate of
  its size and given an extremal obliquity, optimizing the Yarkovsky effect).
  Various classes of solutions, organized into four sectors by the straight
  lines, are discussed in the text. Situation as of December~2014.}  
 \label{fydet}
\end{figure}
\begin{deluxetable}{rlrrrrclrccc}
\tabletypesize{\small}
\tablecaption{List of the Yarkovsky effect detections as of December 2014 \label{yark_det}}
\tablewidth{0pt}
\tablehead{ \multicolumn{2}{c}{Object} & \multicolumn{1}{c}{$\bar r$} &
 \multicolumn{1}{c}{$H$} & \multicolumn{1}{c}{$D$} & \multicolumn{3}{c}{$da/dt$} & 
 SNR & $\cal S$ & Data arc & $N_{\rm rad}$ \\
 & & \multicolumn{1}{c}{(au)} & \multicolumn{1}{c}{(mag)} & \multicolumn{1}{c}{(m)} &
 \multicolumn{3}{c}{($\times 10^{-4}$ au/My)} & & & & }
\startdata
 101955 &     Bennu & 1.10 & 20.6 &  493 & $ -18.95$ &\hspace*{-4mm} $\pm$ &\hspace*{-4mm} $\phantom{3}0.10$ & 194.6 & 1.0 & 1999--2013 & 3 \\ 
   2340 &    Hathor & 0.75 & 20.2 &  210 & $ -17.38$ &\hspace*{-4mm} $\pm$ &\hspace*{-4mm} $\phantom{3}0.70$ &  24.9 & 0.3 & 1976--2014 & 1 \\ 
 152563 &   1992 BF & 0.87 & 19.7 &  510 & $ -11.82$ &\hspace*{-4mm} $\pm$ &\hspace*{-4mm} $\phantom{3}0.56$ &  21.0 & 0.6 & 1953--2011 & 0 \\ 
        &   2009 BD & 1.01 & 28.2 &    4 & $ -489\phantom{.82}$ &\hspace*{-4mm} $\pm$ &\hspace*{-4mm} $35$   &  13.9 & 0.2 & 2009--2011 & 0 \\ 
        & 2005 ES70 & 0.70 & 23.7 &   61 & $ -68.9\phantom{2}$ &\hspace*{-4mm} $\pm$ &\hspace*{-4mm} $\phantom{3}7.9 $ & 8.7 & 0.3 & 2005--2013 & 0 \\ 
   4179 &  Toutatis & 1.96 & 15.1 & 2800 & $ -3.75 $ &\hspace*{-4mm} $\pm$ &\hspace*{-4mm} $\phantom{3}0.45$ & 8.4 & 1.1 & 1934--2014 & 5 \\ 
   2062 &      Aten & 0.95 & 17.1 & 1300 & $ -6.60 $ &\hspace*{-4mm} $\pm$ &\hspace*{-4mm} $\phantom{3}0.80$ & 8.3 & 0.9 & 1955--2014 & 4 \\ 
        &   1999 MN & 0.50 & 21.4 &  175 & $  54.6\phantom{2}$ &\hspace*{-4mm} $\pm$ &\hspace*{-4mm} $\phantom{3}6.8 $ & 8.1 & 0.5 & 1999--2014 & 0 \\ 
   6489 &   Golevka & 2.01 & 19.1 &  280 & $ -4.52 $ &\hspace*{-4mm} $\pm$ &\hspace*{-4mm} $\phantom{3}0.60$ & 7.5 & 0.1 & 1991--2011 & 3 \\ 
   1862 &    Apollo & 1.22 & 16.3 & 1400 & $ -1.58 $ &\hspace*{-4mm} $\pm$ &\hspace*{-4mm} $\phantom{3}0.24$ & 6.5 & 0.2 & 1930--2014 & 2 \\ 
        &   2006 CT & 1.07 & 22.3 &  119 & $ -47.6\phantom{2}$ &\hspace*{-4mm} $\pm$ &\hspace*{-4mm} $\phantom{3}7.7 $ & 6.2 & 0.6 & 1991--2014 & 1 \\ 
   3908 &       Nyx & 1.71 & 17.3 & 1000 & $   9.6\phantom{2}$ &\hspace*{-4mm} $\pm$ &\hspace*{-4mm} $\phantom{3}1.7 $ & 5.8 & 1.1 & 1980--2014 & 2 \\ 
        &  2000 PN8 & 1.22 & 22.1 &  130 & $  49.3\phantom{2}$ &\hspace*{-4mm} $\pm$ &\hspace*{-4mm} $\phantom{3}8.7 $ & 5.7 & 0.7 & 2000--2014 & 0 \\
 162004 &   1991 VE & 0.67 & 18.1 &  827 & $  19.2\phantom{2}$ &\hspace*{-4mm} $\pm$ &\hspace*{-4mm} $\phantom{3}3.6 $ & 5.3 & 0.9 & 1954--2014 & 0 \\ 
  10302 &   1989 ML & 1.26 & 19.4 &  248 & $ 38.7\phantom{2}$ &\hspace*{-4mm} $\pm$ &\hspace*{-4mm} $\phantom{3}7.5 $ & 5.2 & 1.1 & 1989--2012 & 0 \\ 
   2100 & Ra-Shalom & 0.75 & 16.1 & 2240 & $ -5.8\phantom{2}$ &\hspace*{-4mm} $\pm$ &\hspace*{-4mm} $\phantom{3}1.2 $ & 4.7 & 1.0 & 1975--2013 & 4 \\ 
  29075 &   1950 DA & 1.46 & 17.1 & 1300 & $ -2.70 $ &\hspace*{-4mm} $\pm$ &\hspace*{-4mm} $\phantom{3}0.57$ & 4.7 & 0.6 & 1950--2014 & 2 \\ 
  85953 & 1999 FK21 & 0.53 & 18.0 &  590 & $ -11.0\phantom{2}$ &\hspace*{-4mm} $\pm$ &\hspace*{-4mm} $\phantom{3}2.4$ & 4.5 & 0.3 & 1971--2014 & 0 \\ 
 363505 & 2003 UC20 & 0.74 & 18.2 &  765 & $  -4.5\phantom{2}$ &\hspace*{-4mm} $\pm$ &\hspace*{-4mm} $\phantom{3}1.0$ & 4.5 & 0.3 & 1954--2014 & 1 \\ 
        & 2004 KH17 & 0.62 & 21.9 &  197 & $ -42.0\phantom{2}$ &\hspace*{-4mm} $\pm$ &\hspace*{-4mm} $\phantom{3}9.8$ & 4.3 & 0.6 & 2004--2013 & 1 \\ 
  66400 &  1999 LT7 & 0.70 & 19.4 &  411 & $ -35.0\phantom{2}$ &\hspace*{-4mm} $\pm$ &\hspace*{-4mm} $\phantom{3}8.3$ & 4.2 & 0.9 & 1987--2014 & 0 \\ 
        &   1995 CR & 0.45 & 21.7 &  100 & $ -314\phantom{.82}$ &\hspace*{-4mm} $\pm$ &\hspace*{-4mm} $76  $ & 4.2 & 0.8 & 1995--2014 & 0 \\ 
   4034 &    Vishnu & 0.95 & 18.3 &  420 & $ -31.8\phantom{2}$ &\hspace*{-4mm} $\pm$ &\hspace*{-4mm} $\phantom{3}8.0$ & 4.0 & 1.2 & 1986--2014 & 1 \\ 
  85774 & 1998 UT18 & 1.33 & 19.1 &  900 & $ -2.45 $ &\hspace*{-4mm} $\pm$ &\hspace*{-4mm} $\phantom{3}0.63$ & 3.9 & 0.2 & 1989--2014 & 3 \\ 
        &  1994 XL1 & 0.57 & 20.8 &  231 & $ -37.6\phantom{2}$ &\hspace*{-4mm} $\pm$ &\hspace*{-4mm} $\phantom{3}9.8$ & 3.8 & 0.5 & 1994--2011 & 0 \\ 
   3361 &   Orpheus & 1.14 & 19.0 &  348 & $   6.2\phantom{2}$ &\hspace*{-4mm} $\pm$ &\hspace*{-4mm} $\phantom{3}1.7$ & 3.8 & 0.2 & 1982--2014 & 0 \\ 
 377097 &  2002 WQ4 & 1.63 & 19.5 &  422 & $  -9.6\phantom{2}$ &\hspace*{-4mm} $\pm$ &\hspace*{-4mm} $\phantom{3}2.6$ & 3.7 & 0.4 & 1950--2014 & 0 \\ 
 138852 & 2000 WN10 & 0.97 & 20.1 &  328 & $  17.7\phantom{2}$ &\hspace*{-4mm} $\pm$ &\hspace*{-4mm} $\phantom{3}4.9$ & 3.6 & 0.6 & 2000--2014 & 0 \\
 399308 &   1999 GD & 1.07 & 20.8 &  180 & $  47\phantom{.82}$ &\hspace*{-4mm} $\pm$ &\hspace*{-4mm} $13  $ & 3.5 & 0.9 & 1993--2014 & 0 \\
   4581 & Asclepius & 0.96 & 20.7 &  242 & $ -19.7\phantom{2}$ &\hspace*{-4mm} $\pm$ &\hspace*{-4mm} $\phantom{3}5.7$ & 3.5 & 0.4 & 1989--2014 & 1 \\ 
        & 2007 TF68 & 1.36 & 22.7 &  100 & $ -60\phantom{.82}$ &\hspace*{-4mm} $\pm$ &\hspace*{-4mm} $18  $ & 3.4 & 0.7 & 2002--2012 & 0 \\ 
        &   1999 FA & 1.07 & 20.6 &  300 & $ -43\phantom{.82}$ &\hspace*{-4mm} $\pm$ &\hspace*{-4mm} $13  $ & 3.3 & 1.4 & 1978--2008 & 0 \\ 
   2063 &   Bacchus & 1.01 & 17.2 & 1200 & $  -6.6\phantom{2}$ &\hspace*{-4mm} $\pm$ &\hspace*{-4mm} $\phantom{3}2.0$ & 3.2 & 0.8 & 1977--2014 & 2 \\ 
 350462 &  1998 KG3 & 1.15 & 22.2 &  125 & $ -25.2\phantom{2}$ &\hspace*{-4mm} $\pm$ &\hspace*{-4mm} $\phantom{3}7.9$ & 3.2 & 0.4 & 1998--2013 & 0 \\ 
 256004 &   2006 UP & 1.51 & 23.0 &   85 & $ -67\phantom{.82}$ &\hspace*{-4mm} $\pm$ &\hspace*{-4mm} $21   $ & 3.1 & 0.7 & 2002--2014 & 0 \\
  37655 &    Illapa & 0.97 & 17.8 &  950 & $ -10.3\phantom{2}$ &\hspace*{-4mm} $\pm$ &\hspace*{-4mm} $\phantom{3}3.5$ & 3.0 & 0.5 & 1994--2013 & 2 \\ 
\enddata                                                                    
\vspace*{-4mm}
\tablecomments{Reliable detections with SNR larger than 3 are listed: $\bar r = a
 \sqrt{1-e^2}$ is the solar flux-weighted mean heliocentric distance, $H$ is the
 absolute magnitude, $D$ is the diameter derived from the literature when available
 (and obtained here from the EARN Near-Earth Asteroids Database, \httpearn)
 or from absolute magnitude with $15.4$\% albedo, the $da/dt$ and formal uncertainty
 $\sigma_{da/dt}$ are derived from the orbital fit (via $A_2$ and $\sigma_{A_2}$
 values as described in {\em Farnocchia et~al.}, 2013b). ${\rm SNR} = (da/dt)/
 \sigma_{da/dt}$ is the quality of the semimajor axis drift determination, and
 ${\cal S}={\rm SNR}/{\rm SNR}_{\rm max}$, where ${\rm SNR}_{\rm max}$ is the
 maximum estimated ${\rm SNR}$ for the Yarkovsky effect. Data arc indicates the
 time interval over which the astrometric information is available, and
 $N_{\rm rad}$ denotes the number of radar apparitions in the fit.}
\end{deluxetable}
\begin{deluxetable}{rlrrrccc}
\tabletypesize{\small}
\tablecaption{List of the most notable Yarkovsky effect non-detections as of December 2014
 \label{yark_nondet}}
\tablewidth{0pt}
\tablehead{ \multicolumn{2}{c}{Object} & \multicolumn{1}{c}{$\bar r$} &
 \multicolumn{1}{c}{$H$} & \multicolumn{1}{c}{$D$} & $1/{\cal S}$ & Data arc & $N_{\rm rad}$ \\
 & & \multicolumn{1}{c}{(au)} & \multicolumn{1}{c}{(mag)} & \multicolumn{1}{c}{(m)}
 & & & }
\startdata
   3757 &  Anagolay & 1.65 & 19.1 &  390 & 86.8 & 1982--2014 & 1 \\
 247517 & 2002 QY6  & 0.62 & 19.6 &  270 & 56.9 & 2002--2014 & 0 \\
   5797 &     Bivoj & 1.71 & 18.8 &  500 & 53.6 & 1953--2014 & 0 \\
 152742 & 1998 XE12 & 0.62 & 18.9 &  413 & 39.7 & 1995--2014 & 0 \\
   1221 &      Amor & 1.74 & 17.4 & 1100 & 31.0 & 1932--2012 & 0 \\
 225312 & 1996 XB27 & 1.19 & 21.7 &   85 & 20.1 & 1996--2014 & 0 \\
\enddata                                                                    
\vspace*{-4mm}
\tablecomments{Notable non-detections of the Yarkovsky effect with
 $1/{\cal S}>10$ are listed. Columns as in Table~\ref{yark_det}.}
\end{deluxetable}

Figure~\ref{fydet} depicts the distribution of NEAs in the SNR and SNR$_{\rm max}$
space that we divide into four regions:
\begin{itemize}
\item We consider cases with ${\rm SNR}>3$ and ${\cal S} < 1.5$ to be {\em valid detections}
 because the estimated value is no more than $50$\% larger than expected, perhaps
 as a result of unusually low density or a size far smaller than assumed. 
 Table~\ref{yark_det} lists the 36 objects with valid Yarkovsky detections
 given currently available astrometry.
\item {\em Spurious detections} are those with ${\rm SNR}>3$ and ${\cal S} > 1.5$. Many
 of these are due to astrometric errors in isolated observation sets, such as
 precoveries, and can be moved to the left in Fig.~\ref{fydet} by deweighting the
 questionable data. We find 56 cases in this category, but only 12 with SNR $>4$.
 There are two spurious cases with ${\rm SNR}>10$ and ${\cal S} \gtrsim 10$ that cannot
 be due to astrometric errors and are yet unlikely to be attributed to the
 Yarkovsky effect. 
\item There are a number of objects with relatively low values for $\sigma_{A_2}$
 and yet the orbit does not reveal an ${\rm SNR}>3$ detection (denoted as {\em weak
 signal} zone on Fig.~\ref{fydet}). Specifically, these
 cases have SNR$_{\rm max}>3$ and ${\rm SNR}<3$, with ${\cal S} < 2/3$. These cases are
 potentially interesting because they generally indicate a mid-range obliquity
 and, despite the lack of significance in the $A_2$ estimate, useful bounds can
 be still placed on the Yarkovsky mobility of the object. We find 35 such cases
 in the current NEA catalog, six of which have ${\cal S} < 0.05$ (Table~\ref{yark_nondet}).
 In fact, this class warrants further dedicated analysis, similar to the search of
 new detections.
\item The vast majority of NEAs are currently uninteresting due to ${\rm SNR}<3$ and
 SNR$_{\rm max}<3$ meaning that no detection was found nor was one reasonably expected.
\end{itemize}

It is worth noting that objects with non-principal axis rotation states can reveal 
the Yarkovsky effect (e.g., {\em Vokrouhlick\'y et~al.}, 2005a); (4179) Toutatis is a 
large, slowly tumbling asteroid (e.g., {\em Hudson and Ostro}, 1995) with Yarkovsky 
$\mathrm{SNR}\simeq 8$ (and ${\cal S}\simeq 1$) due to an extensive set of radar ranging 
data. Also the much smaller asteroid (99942) Apophis, which has been reported to 
have a measurable polar precession ({\em Pravec et~al.}, 2014), presently has a 
solid Yarkovsky signal with $\mathrm{SNR}\simeq 1.8$ (and ${\cal S}<1$), though not high
enough to be listed in Table~\ref{yark_det}, but still significant in light of 
the abundant radar astrometry available for Apophis ({\em Vokrouhlick\'y et~al.},
2015). Similarly, binary asteroid systems may also reveal Yarkovsky drift in their 
heliocentric orbits (e.g., {\em Vokrouhlick\'y et~al.}, 2005b), although none presently 
appears in Table~\ref{yark_det}. We note that (363599) 2004~FG11 has a satellite 
({\em Taylor et~al.}, 2012) and currently has a Yarkovsky $\mathrm{SNR}\simeq 2.8$ 
(and ${\cal S}\simeq 1$). 

\bigskip
\noindent
\textbf{3.2 YORP effect}
\bigskip

Analyses of small-asteroid populations indicate clear traits of their
evolution due to the YORP effect, both in rotation-rate and
obliquity (Secs.~4.5, 5.1 and 5.2). Accurate observations
of individual objects, however, presently do not permit detection of
the secular change in obliquity and reveal only the secular
effect in rotation rate. Even that is a challenging task because the
YORP torque has a weak effect on kilometer-sized asteroids at roughly 1~au 
heliocentric distance. Similar to the case of the Yarkovsky effect, the
YORP detection is enabled via accurate measurement of a phase $\phi$
associated with the rotation rate. This is because when the rotation frequency 
$\omega$ changes linearly with time, $\omega(t)=\omega_0+(d\omega/dt)\,t$
(adopting the simplest possible assumption, since $d\omega/dt$ may have
its own time variability), the related phase $\phi$ grows quadratically 
in time, $\phi(t)=\phi_0+\omega_0\,t+\case12 (d\omega/dt)\,t^2$. Additionally, 
other perturbations (such as an unresolved weak tumbling) do not produce an
aliasing signal that would disqualify YORP detection. So the determination 
of the YORP-induced change in the rotation rate $d\omega/dt$ may basically 
alias with the rotation rate frequency $\omega_0$ itself in the $d\omega/dt=0$ 
model. This is because small variations in $\omega_0$ propagate linearly in
time in the rotation phase. The YORP detection stems from the ability to
discern this linear trend due to the $\omega_0$ optimization and the 
quadratic signal due to a non-zero $d\omega/dt$ value. In an ideal situation
of observations sufficiently densely and evenly distributed over a
given time interval $T$, one avoids the $\omega_0$ and $d\omega/dt$ correlation
setting time origin at the center of the interval. At the interval limits
the YORP effect manifests via phase change $\simeq \case18 (d\omega/dt)\,T^2$.
Therefore, a useful approximate rule is that the YORP effect is
detected when this value is larger than the phase uncertainty $\delta\phi$
in the observations. Assuming optimistically $\delta\phi\simeq 5^\circ$
and $T$ about a decade, the limiting detectable $d\omega/dt$ value is
$\simeq 5\times 10^{-8}$~rad/d$^2$. Obviously, detection favors longer
time-base $T$ if accuracy of the early observations permits. In practice,
late 1960s or early 1970s was the time when photoelectric photometry was 
introduced and allowed reliable-enough lightcurve observations. This sets
maximum $T$ of about 40 years today for bright-enough objects (e.g.,
(1620) Geographos, {\em \v{D}urech et~al.}, 2008a); see, for completeness,
an interesting YORP study for asteroid (433) Eros by {\em \v{D}urech} (2005). 
We should also mention 
that $\omega$ and $\phi$ above denote sidereal rotation rate and phase,
respectively. Hence to convert asteroid photometry to $\phi$ one needs
to know orientation of its spin axis in the inertial space and the shape 
model. Their solution may increase the realistic
uncertainty in $d\omega/dt$ if compared to the simple estimate
discussed above.
\begin{deluxetable}{rlrclcrccll}
\tabletypesize{\small}
\tablecaption{List of the YORP effect detections as of September 2014 \label{yorp_det}}
\tablewidth{0pt}
\tablehead{ \multicolumn{2}{c}{Object} & \multicolumn{3}{c}{$d\omega/dt$} &
 \multicolumn{1}{c}{$H$} & \multicolumn{1}{c}{$P$} & $\gamma$ & $\bar{r}$ & Reference \\
 & & \multicolumn{3}{c}{($\times 10^{-8}$ rad/d$^2$)} & \multicolumn{1}{c}{(mag)} &
 \multicolumn{1}{c}{(h)} & (deg) & (au) & }
\startdata
 54509 & YORP       & $350\phantom{.5}$ &\hspace*{-4mm} $\pm$ &\hspace*{-4mm} $35$ & 22.6 & $\phantom{1}0.203$ & $173$
  & 0.98 &  {\em Lowry et~al.} (2007); {\em Taylor et~al.} (2007) \\
 25143 & Itokawa    & $3.5$ &\hspace*{-4mm} $\pm$ &\hspace*{-4mm} $\phantom{3}0.4$ & 18.9 & $12.132$           & $178$
  & 1.27 & {\em Lowry et~al.} (2014) \\
  1620 & Geographos & $1.2$ &\hspace*{-4mm} $\pm$ &\hspace*{-4mm} $\phantom{3}0.2$ & 15.6 & $\phantom{1}5.223$ & $152$
  & 1.18 & {\em \v{D}urech et~al.} (2008a) \\
  1862 & Apollo     & $5.5$ &\hspace*{-4mm} $\pm$ &\hspace*{-4mm} $\phantom{3}1.2$ & 16.3 & $\phantom{1}3.065$ & $162$
  & 1.22 & {\em Kaasalainen et~al.} (2007); {\em \v{D}urech et~al.} (2008b) \\
  3103 & Eger       & $1.4$ &\hspace*{-4mm} $\pm$ &\hspace*{-4mm} $\phantom{3}0.6$ & 15.3 & $\phantom{1}5.710$ & $176$
  & 1.32 & {\em \v{D}urech et~al.} (2012) \\
  1865 & Cerberus   & $< 0.8$ &\hspace*{-4mm} &\hspace*{-4mm} & 16.8 & $\phantom{1}6.803$
  & $178$ & 0.96 &  {\em \v{D}urech et~al.} (2012) \\
\enddata
\vspace*{-4mm}
\tablecomments{For each of the asteroids with the YORP effect detected we give: (i) rotation rate
 change $d\omega/dt$ derived from the photometric data, (ii) absolute magnitude $H$, (iii)
 rotation period $P$, (iv) obliquity $\gamma$, and (v) the solar flux weighted mean heliocentric
 distance $\bar{r}= a\,\sqrt{1-e^2}$, with semimajor axis $a$ and eccentricity $e$.
 In the case of (1865)
 Cerberus, the observational limit $|d\omega/dt|< 0.8\times 10^{-8}$~rad/d$^2$ is non-trivial for
 a body of its size, orbit and rotation state. Less severe limits on $|d\omega/dt|$ were also
 derived for (2100) Ra-Shalom ({\em \v{D}urech et~al.}, 2012) and (433) Eros ({\em \v{D}urech},
 2005).}

\end{deluxetable}
\begin{figure}[t]
 \epsscale{1.}
 \plotone{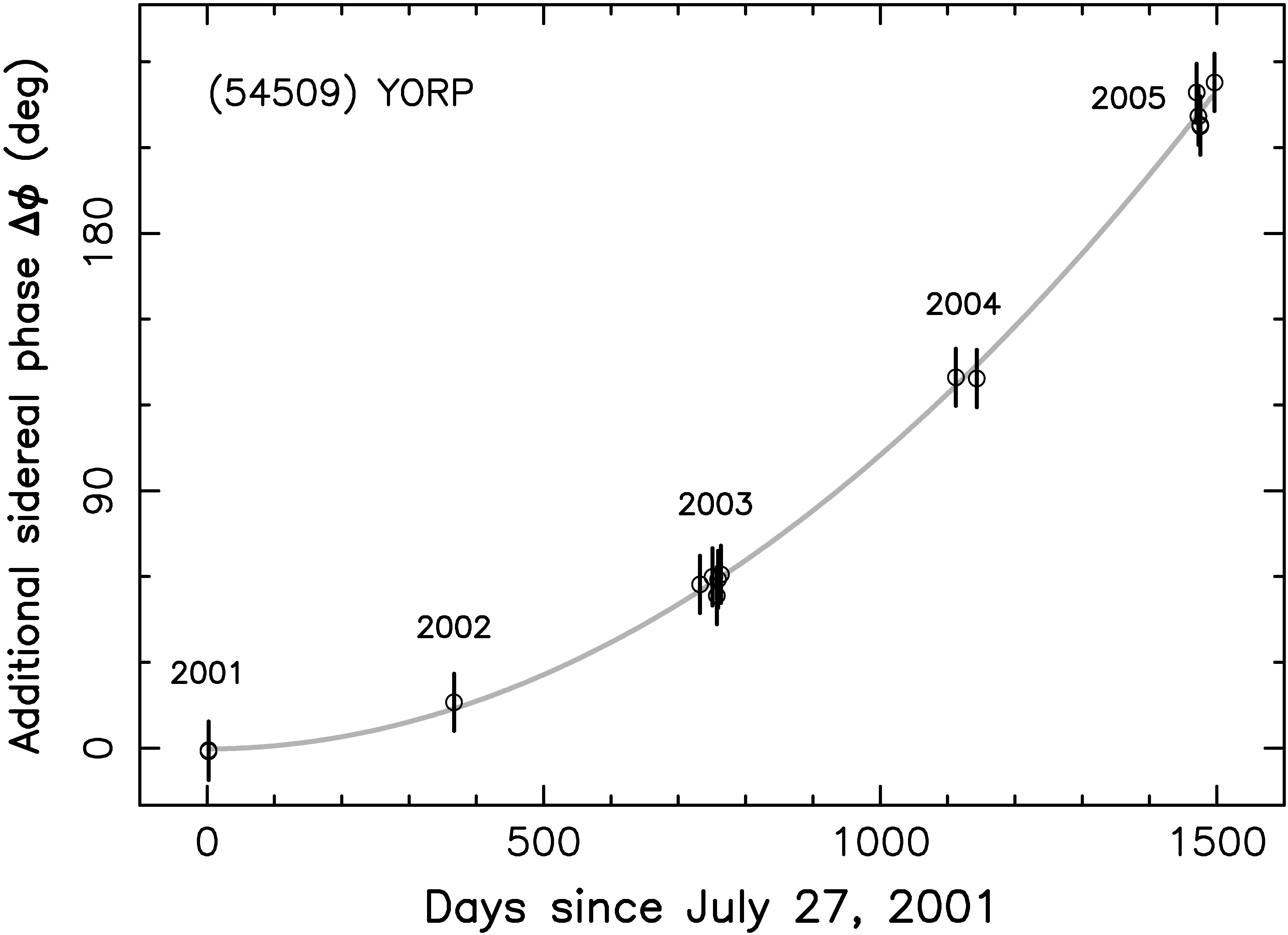}
 \caption{\small Advance of the sidereal rotation phase $\Delta \phi$
  (ordinate in degrees) vs time (in days) for a small Earth-coorbital
  asteroid (54509)~YORP. Symbols are measurements with their estimated
  uncertainty, as follow from assembling the radar observations at
  different apparitions. The gray line is a quadratic progression $\Delta \phi
  = \case12 (d\omega/dt) t^2$, with $d\omega/dt=350\times 
  10^{-8}$~rad/d$^2$. Time origin set arbitrarily to July~27, 2001
  corresponding to the first measurement. Adapted from {\em Taylor et~al.}
  (2007).}  
 \label{fyo}
\end{figure}

Figure~\ref{fyo} shows an example of detected quadratic advance in
sidereal rotation phase $\phi$ in the case of a small coorbital asteroid
(54509)~YORP (see {\em Lowry et~al.}, 2007; {\em Taylor et~al.}, 2007). 
The expected YORP value of rotation-rate change matched the observed value,
thus allowing interpretion of the signal as a YORP effect detection,
though an accurate comparison is prohibited by lack of knowledge of the 
full shape of this body (due to repeated similar viewing geometry from
the Earth). A complete list of the YORP detections, as of the publication
date, is given in Table~\ref{yorp_det}. To appreciate their accuracy we 
note that they correspond to a tiny change in sidereal rotation 
period by a few milliseconds per year: $1.25$~ms/y for (54509)~YORP 
to a maximum value of $45$~ms/y for (25143)~Itokawa. While not numerous
at the moment, we expect the list will more than double during the next 
decade. There are presently two asteroids, (1620)~Geographos and (1862)~Apollo, 
for which both Yarkovsky and YORP effects have been detected. These
cases are of special value provided an accurate-enough physical model
of the body is available (see {\em Rozitis et~al.}, 2013; {\em Rozitis
and Green}, 2014).

(25143) Itokawa holds a special place among the asteroids with the YORP effect
detected. Not only was this the first asteroid for which YORP detection
was predicted ({\em Vokrouhlick\'y et~al.}, 2004), but the shape of this body 
is known very accurately thanks to
the visit of the Hayabusha spacecraft. This has led researchers to push the attempts
for an accurate YORP prediction to an extreme level (e.g., {\em Scheeres et~al.},
2007; {\em Breiter et~al.}, 2009; {\em Lowry et~al.}, 2014), realizing that
the results depend in this case very sensitively on the small-scale irregularities 
of the shape (see {\em Statler} 2009 for a general concept). However, in spite of
an uncertainty in the YORP prediction, the most detailed computation 
consistently predicted deceleration of the rotation rate by YORP, as opposed
to the detected value (Table~\ref{yorp_det}). A solution to this conundrum
has been suggested by {\em Scheeres et~al.} (2007), who proposed that the
difference in density between the ``head'' and ``body'' of this asteroid may
shift significantly the center-of-mass. This effect introduces an extra
torque component which could overrun the YORP torque, canonically
computed for homogeneous bodies, and make the predicted deceleration become
acceleration of the rotation rate. {\em Lowry et~al.} (2014) adopted this
solution, predicting that the two parts of Itokawa have a very different
densities of $\simeq 1.75$~g/cm$^3$ and $\simeq 2.85$~g/cm$^3$. Nevertheless,
the situation may be still more complicated: {\em Golubov and Krugly}
(2012) have shown that transverse heat communication across boulder-scale
features on the surface of asteroids may cause a systematic trend toward
acceleration of the rotation rate. Indeed, in the most complete works so
far, {\em Golubov et~al.} (2014) and {\em \v{S}eve\v{c}ek et~al.} (2015),
show that the detected acceleration
of Itokawa's rotation rate may be in large part due to detailed modeling
of the effects described by {\em Golubov and Krugly} without invoking large
density difference in the asteroid. The complicated case of Itokawa thus
keeps motivating detailed modeling efforts of the YORP effect.
Luckily, not all asteroidal shapes show such an extreme sensitivity on
the small-scale surface features (e.g., {\em Kaasalainen and Nortunen}, 2013),
allowing thus an easier comparison between the detected and predicted
YORP signals.

On a more general level, we note that in spite of rotation periods 
ranging from a fraction of an hour to more than $12$~hours, all five 
asteroids for which the YORP effect was detected reveal
acceleration of the rotation rate. It is not known yet, whether this
expresses observational bias against detection of the YORP-induced
deceleration of the rotation rate, or whether it points toward the true
asymmetry in YORP's ability to accelerate vs decelerate rotation rate.
Note that one would statistically expect to detect YORP deceleration of
the rotation rate principally among asteroids rotating slowly, but this
is exactly where accurate photometric observations are especially
difficult. Efforts with the goal to detect the YORP effect for asteroids with 
rotation periods in the $20$ to $40$~hr range are under way, with the
results expected in the next couple of years. Hopefully, they will help in
setting the issue of possible asymmetry in the YORP effect on $\omega$.

\bigskip
\noindent
 \textbf{3.3 BYORP effect}
\bigskip

The BYORP effect has not been directly observed as of yet, although 
some predictions stemming from this effect have been confirmed. There are 
currently significant campaigns observing binary asteroids 
to search for predicted outcomes of the BYORP effect, both in isolation 
or mixing with other evolutionary effects. The basic technique for 
detecting the BYORP effect as it acts in solitude was proposed by {\em 
McMahon and Scheeres} (2010b) and suggests that computing the drift in 
a binary system's mean anomaly due to changes in the semi-major axis is 
the most effective approach, as this drift will increase quadratically 
in time as compared to purely Keplerian motion. The relative change 
$\Delta M$ in the mean anomaly $M$ of a binary asteroid due to the BYORP 
effect in time $t$ is: $a\Delta M  =  - \frac{3}{4} \,n\,(da/dt)\, t^2$,
where $n$ is the binary mean motion and $(da/dt)$ should be substituted 
from Eq.~(\ref{eq:BYORP_a}). The corresponding delay, or advance, in 
occultation timing of the binary is $\simeq - \frac{3}{4} [(da/dt)/a]\,
t^2$.

{\em McMahon and Scheeres} (2010b) provide a table of known and possibly 
synchronous binary asteroids along with an estimate of mean anomaly 
drift, based on scaling the computed (66391) 1999~KW4 BYORP coefficient to 
the  different asteroid systems, accounting for secondary size, system mass, 
and heliocentric orbit. As these stated drifts make a strong assumption 
in applying the KW4 BYORP coefficient they are not true predictions, but 
rather provide a prediction of relative strength of the BYORP effect 
for different bodies. Petr Pravec has expanded this list of predicted 
drift rates, making them accessible in the Binary Asteroid Database 
(\httpbin) and indicating which should have the largest, and hence easiest 
to detect, drifts along with other information of use to observers. 

This list represents an active longer-term campaign by Pravec and 
colleagues to observe binary asteroid systems during predicted occultation 
events. The most significant result of this effort to
date has been focused on the binary asteroid (175706) 1996~FG3
({\em Scheirich et~al.}, 2015). For this body, observations over a $17$~y 
time span provided a strong ``zero'' constraint on the BYORP drift rate. 
While not a direct detection of the BYORP effect, this is fully consistent 
with a current prediction that involves the BYORP-tide equilibrium state. 
The confirmation of a binary system in this state has scientific implications 
as it means that the tidal dissipation that occurs within a rapidly spinning 
primary body can be determined once the BYORP coefficient for a secondary 
asteroid is determined. Although it cannot be directly measured when in 
such an equilibrium, it is possible to estimate the BYORP coefficient 
based on detailed models of the secondary and its albedo, such as could 
be obtained by an in situ spacecraft. Thus, a space mission to a binary in 
such a state could provide an unprecedented view into the internal geophysics 
of a rubble pile. 
\begin{figure}[t]
 \epsscale{1.}
 \plotone{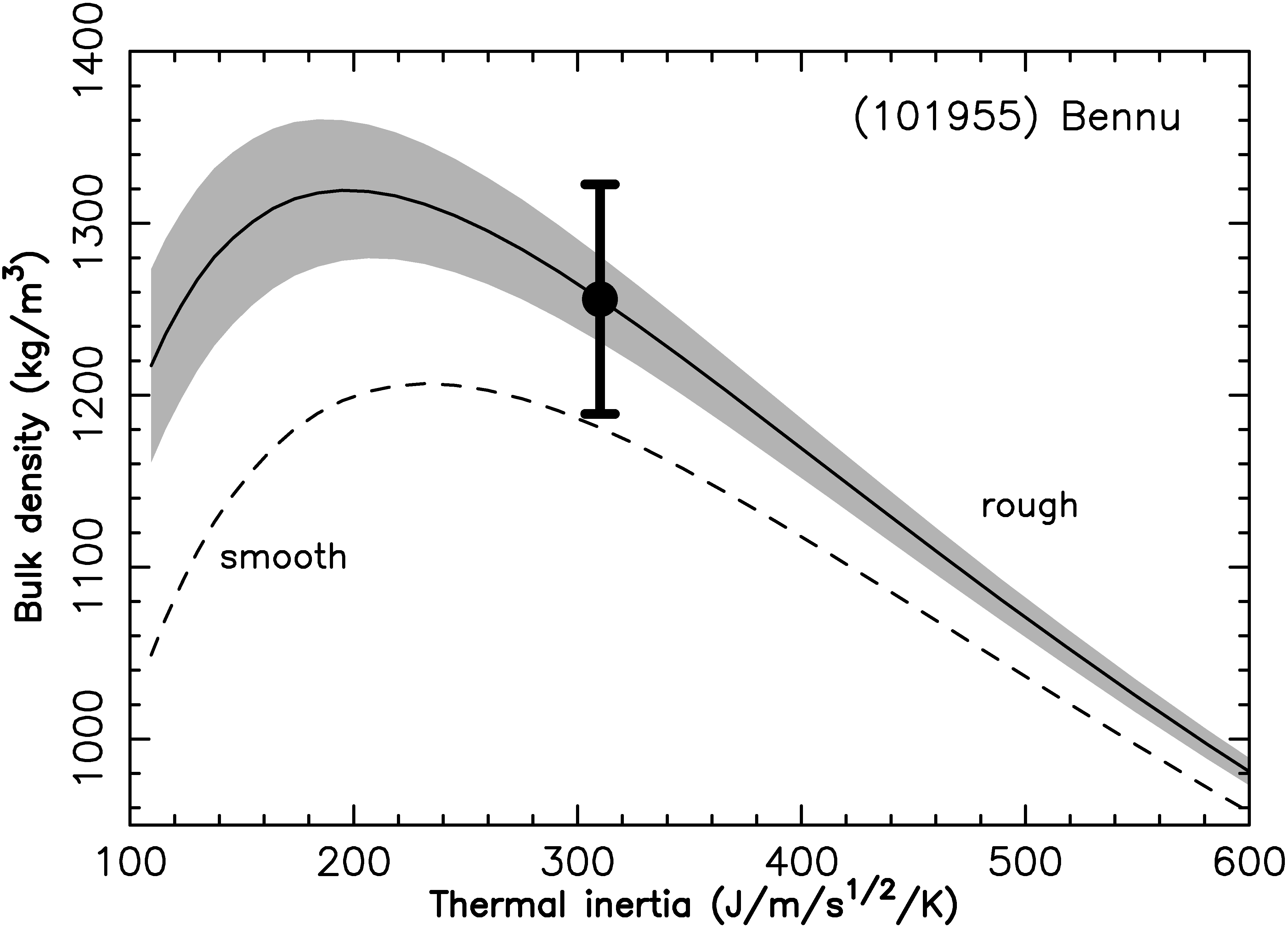}
 \caption{\small Bulk density $\rho$ solution for (101955) Bennu from detected
  value of the Yarkovsky orbital effect as a function of the surface thermal
  inertia $\Gamma$. The dashed line corresponds to the $da/dt=$~const.
  solution for a smooth-surface model, taking into account a detailed shape model
  and a nonlinear boundary condition. The solid line accounts for $50$\%
  small-scale roughness in each of the surface facets of the shape model,
  while the gray zone takes into account the estimated $\sim 17$\% uncertainty
  in the roughness value. The non-linearity of the $da/dt$ isoline in the $\rho$ vs
  $\Gamma$ plane follows from Eqs.~(\ref{yark_ana1}) and (\ref{yark_ana3}).
  Adapted from {\em Chesley et~al.} (2014).}  
 \label{fbennu}
\end{figure}

Other bodies of current interest include any binary systems with a 
synchronous secondary. A direct detection of BYORP is feasible if the 
body is in an expansive state, although the relatively short lifetime 
predicted for such binaries would imply that finding such a binary may 
be difficult. Similarly for a contractive state, as this should be heading 
towards a BYORP-tide equilibrium. Additional measurements are important, 
however, as the number of binaries found to be in the equilibrium state 
relative to the number found in expansive or contractive states will be an 
important measurement with implications beyond the BYORP effect in isolation. 
Specifically, such observations could provide insights into the internal 
tidal dissipation of energy that occurs for rubble pile binary asteroid 
primaries (e.g., {\em Jacobson and Scheeres}, 2011b). 
\bigskip

\centerline{\textbf{4. APPLICATIONS OF THE YARKOVSKY EFFECT}}
\bigskip

\bigskip
\noindent
 \textbf{4.1 Physical properties of asteroids}
\bigskip

The Yarkovsky effect can be used as a tool to probe the nature of individual 
asteroids. This is possible because an asteroid's Yarkovsky drift is a 
manifestation of several of its physical properties, and so a direct 
measurement of $da/dt$ allows insight into the characteristics of the 
body. Of primary importance are the obliquity, size and mass of the 
asteroid, although the thermal and reflective properties and the rotation 
rate are also important.

Not surprisingly, the more that is known about the asteroid the more that 
can be divined from a Yarkovsky detection. In the weakest situation, which 
is not so unusual, we have only $da/dt$ and the absolute magnitude $H$. Even 
in this case we can already put meaningful constraints on the obliquity of 
the body through the $\cos\gamma$ dependence. For instance, the sign of 
$da/dt$ reveals immediately whether the rotation is retrograde or direct. 
Moreover, the value of $\cal S$ from Table~\ref{yark_det} can serve as a 
proxy for $|\cos\gamma|$, while variations in $\rho D$ and $\Theta_\omega$ add
uncertainty to this estimate. {\em Vokrouhlick\'y et~al.} (2008) used this 
principle to infer that (152563) 1992~BF must have obliquity $\gamma > 
120^\circ$, after accounting for reasonable variations in other unknowns.

If the spin state of the body is known, generally from some 
combination of radar imaging and optical light curves, we have a much 
clearer insight into the nature of the body because $\cos\gamma$ is 
removed as an unknown and the thermal parameter $\Theta_\omega$ is better 
constrained. Indeed, in such cases we are left with a simple 
relationship between $\rho D$ and the thermal inertia $\Gamma$. But 
the diameter $D$ can be measured directly by radar, or inferred 
from taxonomic type or measured albedo, or can just be derived from 
an assumed distribution of asteroid albedo, allowing the constraint to be
cast in terms of the bulk density $\rho$ and thermal inertia $\Gamma$. 
The gray region of Fig.~\ref{fbennu} depicts this type of constraint 
for the case of (101955)~Bennu. The peak in $\rho$ seen in 
Fig.~\ref{fbennu} is associated with $\Theta_\omega \simeq 1$ where the 
Yarkovsky effect obtains it maximum effectiveness. This characteristic 
peak in the $\rho$ vs. $\Gamma$ relationship often allows strict upper 
bounds on $\rho$ (e.g., {\em Chesley et~al.}, 2003).

We note that the degeneracy between $\rho$ and $\Gamma$ could in principle be 
broken by an independent estimate of $\rho$ that would allow a direct estimate 
of $\Gamma$, albeit with the possibility of two solutions. While this approach 
has so far not been possible, we anticipate it here as a natural outcome of 
the first detection of the Yarkovsky effect on a well-observed binary system.

Another approach to breaking the correlation between $\rho$ and $\Gamma$ 
makes use of measurable solar radiation pressure deviations on the orbit, 
which yields an area-to-mass ratio. With a size estimate, an independent
mass estimate can lead to a double solution for the thermal inertia of the 
body (e.g., {\em Mommert et~al.}, 2014).

The alternative approach has been applied successfully in a few special cases 
to date. Specifically, observations of an asteroid's thermal emissions can 
afford independent constraints on the thermal inertia, breaking the degeneracy
between $\rho$ and $\Gamma$, allowing a direct estimate of the asteroid's 
bulk density. Perhaps the most striking example here is the case of 
(101955)~Bennu, which has a well-constrained shape, spin state and thermal 
inertia. When these are linked with the high precision $da/dt$ estimate 
(Table~\ref{yark_det}) the result is a bulk density of $1260\pm 70\;
\mathrm{kg/m}^3$ (Fig.~\ref{fbennu}), where the formal precision is better 
than 6\% ({\em Chesley et~al.}, 2014). Other similar cases include (1862)~Apollo, 
(1620)~Geographos and (29075) 1950~DA (respectively, {\em Rozitis et al.}, 2013,
2014; {\em Rozitis and Green}, 2014). In each of these cases the authors 
combine $da/dt$, radar imaging and thermal measurements to derive the bulk 
density of the asteroid.

In the best cases of Yarkovky detection, where we also have a shape model, 
spin state and thermophysical characterization, one can infer the local gravity 
of the body. This can be of profound engineering interest for the asteroid 
targets of space missions, e.g., (101955)~Bennu. The mission design challenges 
for the OSIRIS-REx mission are significantly eased due to the Yarkovsky 
constraint on Bennu's mass and bulk density. Another such case is (29075)
1950~DA, which is not a space mission target, and yet the estimates of local 
surface gravity derived from Yarkovsky have profound implications. {\em Rozitis 
et~al.} (2014) found that their thermal measurements, when combined with 
the Yarkovsky drift reported for 1950~DA by {\em Farnocchia and Chesley} (2014),
required a low asteroid mass. The estimated mass was so low, in fact, that it 
implied that the equatorial surface material on 1950~DA is in tension due to 
centrifugal forces. And yet the estimated thermal inertia was low enough 
that it required a loose, fine-grained regolith on the surface. This seeming 
contradiction is most readily resolved by the action of cohesive forces due 
to van der Waals attraction between regolith grains, and represents the first 
confirmation of such forces acting on an asteroid, which were already
anticipated by {\em Scheeres et~al.} (2010). And so, through a curious
interdisciplinary pathway, the measurement of the Yarkovsky drift on 1950~DA 
reveals the nature of minute attractive forces at work in the asteroid's regolith.
\smallskip

\noindent{\bf Population implications.-- }The 
discussion above treats Yarkovsky detections in a case-by-case manner, 
deriving additional information for the specific asteroid at hand. However, 
the wealth of Yarkovsky detections listed in Table~\ref{yark_det} allows an insight
into the near-Earth asteroid population as a whole. Of particular interest is 
the distribution of obliquities implied by the tabulated detections, of which 
$28$ out of $36$ detections reveal $da/dt < 0$ and thus about 78\% of the sample 
requires retrograde rotation (see also Fig.~\ref{fg}).

This excess of retrograde rotators represents an independent confirmation of 
a result first reported by {\em La~Spina et~al.} (2004). The mechanism for an 
excess of retrograde rotators in the near-Earth population is a result of the 
Yarkovsky-driven transport mechanism (e.g., {\em Morbidelli and Vokrouhlick\'y},
2003). The location of the $\nu_6$ resonance at the inner edge of the main belt 
implies that main belt asteroids entering the inner solar system through this 
pathway must have $da/dt < 0$ and thus retrograde rotation. Direct rotators will 
tend to drift away from the resonance. Asteroids entering the inner solar 
system through other resonance pathways, principally the 3/1 mean motion resonance with
Jupiter, may drift either in or out into the resonance, and so will have parity 
between retrograde and direct rotators. {\em Farnocchia et~al.} (2013b) analyze this 
retrograde prevalence, including selection effects among the Yarkovsky detections, 
and find that it is fully consistent with the Yarkovsky-driven transport, and point 
out that this can be used to derive a distribution of the obliquities of NEAs.

\bigskip
\noindent
 \textbf{4.2 Impact hazard assessment}
\bigskip

Most reported potential impacts are associated with newly discovered objects 
for which the uncertainty at the threatening Earth encounter is dominated by 
the uncertainties in the available astrometric observations. However, as the 
astrometric dataset grows, the fidelity of the force model used to propagate 
the asteroid from discovery to potential impact becomes more and more important. 
For a few asteroids with extraordinarily precise orbits, the Yarkovsky effect 
is a crucial aspect of an analysis of the risk posed by potential impacts on 
Earth. When the Yarkovsky effect is directly revealed by the astrometric data 
the analysis approach is straightforward as is the case for (101955) Bennu and 
(29075) 1950~DA (e.g., {\em Milani et~al.}, 2009; {\em Chesley et~al.}, 2014; 
{\em Farnocchia and Chesley}, 2014).

However, there are some cases where the astrometry provides little or no 
constraint on the the Yarkovsky effect and yet Yarkovsky drift is a major 
contributor to uncertainties at a potentially threatening Earth encounter. 
In these situations we are forced to assume distributions on albedo, 
obliquity, thermal inertia, etc., and from these we can derive
a distribution of $A_2$ or $da/dt$. A Monte Carlo approach with these 
distributions allows us to better represent uncertainties at the threatening 
Earth encounter, and thereby compute more realistic impact probabilities. 
This technique has been necessary for (99942)~Apophis and has been applied
by {\em Farnocchia et~al.} (2013a) before {\em Vokrouhlick\'y et~al.} (2015)
made use of rotation state determination of this asteroid. See {\em Farnocchia 
et~al.} (this volume) for a more complete discussion of Yarkovsky-driven 
impact hazard analyses.

\bigskip
\noindent
 \textbf{4.3 Meteorite transport issues}
\bigskip

The Yarkovsky effect, with its ability to secularly change the semimajor
axes of meteoroids (precursors of meteorites, which are believed to
be fragments of larger asteroids located in the main belt between the
orbits of Mars and Jupiter), was originally proposed to be the main
element driving meteorites to the Earth (see already {\em \"Opik}, 1951
or {\em Peterson}, 1976). However, direct transport from the main belt,
say as a small body slowly spiraling inward toward the Sun by the
Yarkovsky effect, required very long timescales and unrealistic
values of the thermal parameters and/or rotation rates for meter size bodies.
Moreover, AM/PM fall statistics and measured pre-atmospheric trajectories
in rare cases (like the P\v{r}{\'\i}bram meteorite) indicated many meteorites
had orbits with the semimajor axis still close to the main belt values.

The problem has been overcome in the late 1970s and early 1980s by
advances in our understanding of asteroid dynamics. Numerous works
have shown that the transport routes that connect main belt objects to
planet-crossing orbits are in fact secular and mean motion resonances with
giant planets, such as the $\nu_6$ secular resonance at the lower
border of the main asteroid belt and/or the 3/1 mean motion resonance
with Jupiter. Putting this information together with the Yarkovsky
effect, {\em Vokrouhlick\'y and Farinella} (2000) were able to
construct a model in which meteoroids or their immediate precursor
objects are collisionally born in the inner and/or central parts of the main
belt, from where they are transported to the resonances by the Yarkovsky effect.
En route, some of the precursors may fragment, which can
produce new swarms of daughter meteoroids which eventually reach the
escape routes to planet crossing orbits. With this model, {\em Vokrouhlick\'y
and Farinella} could explain the distribution of the cosmic-ray exposure 
ages of stony meteorites as a combination of several timescales: (i) 
the time it take a meteoroid to collisionally break, (ii) the 
time it takes a meteoroid to travel to a resonance, (iii) the time it takes
for that resonance to deliver the meteoroid to an Earth-crossing orbit, and 
(iv) the time it takes the meteoroid on a planet-crossing orbit to hit 
the Earth.

While successful to the first order, this model certainly contains
a number of assumptions and potentially weak elements, especially
in the light of subsequent rapid development of the YORP effect theory,
that warrant further work. For instance, one of the difficulties in 
refining the meteorite delivery models is the uncertainty in identification 
of the ultimate  parent asteroid (or asteroids) for a given meteorite class
(e.g., {\em Vernazza et~al.}, this volume). Thus, among the ordinary 
chondrites we have a reasonable guess that LL-chondrites originate from 
the Flora region (or the asteroid (8)~Flora itself) and the L-chondrites 
originate from disruption of the Gefion family. There were
numerous guesses for the H-chondrite source region (such as
the asteroid (6)~Hebe), but none of them has been unambiguously 
confirmed. The model presented by {\em Nesvorn\'y et~al.} (2009),
while more educated in the choice of the L-chondrite source
region than the previous work of {\em Vokrouhlick\'y and Farinella} 
(2000), requires immediate parent bodies of these meteorites,
$5-50$~m in size, to reach
the powerful 3/1 mean motion resonance with Jupiter. This means they 
should have migrated by the Yarkovsky effect some $0.25-0.3$~au
from their source location in less than half a billion years.
While this is not a problem in a scenario where the bodies
rotate about the body-fixed axis whose direction is preserved
in the inertial space, it is not clear if this holds when
the bodies would start to tumble or their axes to evolve rapidly 
due to the YORP effect. Clearly, more work is needed
to understand the Yarkovsky effect in the small-size limit for
bodies whose spin axis may undergo fast evolution.

\bigskip
\noindent
 \textbf{4.4 Orbital convergence in asteroid families and pairs}
\bigskip

Over the past decade the Yarkovsky and YORP effects have helped to
significantly boost our knowledge of the asteroid families (e.g.,
{\em Nesvorn\'y et~al.}, this volume). This is because they represent
a unique time-dependent process in modeling their structure, 
allowing thus for the first time to constrain their ages.

The most accurate results are obtained for young-enough families
(ages $<10$~m.y., say), for which effects of the deterministic chaos
are weak. As shown in the pioneering works of {\em Nesvorn\'y et~al.}
(2002, 2003), the basic tool to determine the origin of the family is
provided by the convergence of orbital secular angles (the nodal and
pericenter longitudes $\Omega$ and $\varpi$) at some moment in the past. 
Because the rate at which these angles precess in space depends 
sensitively on the semimajor axis value, the past values of
$\Omega$ and $\varpi$ of the family members depend on their Yarkovsky 
drift-rates $da/dt$. This contribution may not be negligible,
because the changes in precession rates produce effects that grow
quadratically in time (the same way as described in Secs.~3.1 and 
3.2 for longitude in orbit or sidereal rotation phase). Thus 
{\em Nesvorn\'y and Bottke} (2004) were able to significantly improve
the uncertainty in the age of the Karin family by including the 
Yarkovsky effect in their model. At the same time, this work provided
an effective
detection of the Yarkovsky effect for the main belt asteroids.
This technique has been later used for age constrains of several
other young families (e.g., {\em Novakovi\'{c}}, 2010; {\em Novakovi\'{c}
et~al.} 2012, 2014), including sub-m.y. old clusters (e.g., {\em Nesvorn\'y et~al.},
2006; {\em Nesvorn\'y and Vokrouhlick\'y}, 2006; {\em Nesvorn\'y et~al.}, 2008;
{\em Vokrouhlick\'y et~al.}, 2009).
\begin{figure*}[t]
 \epsscale{1.8}
 \plotone{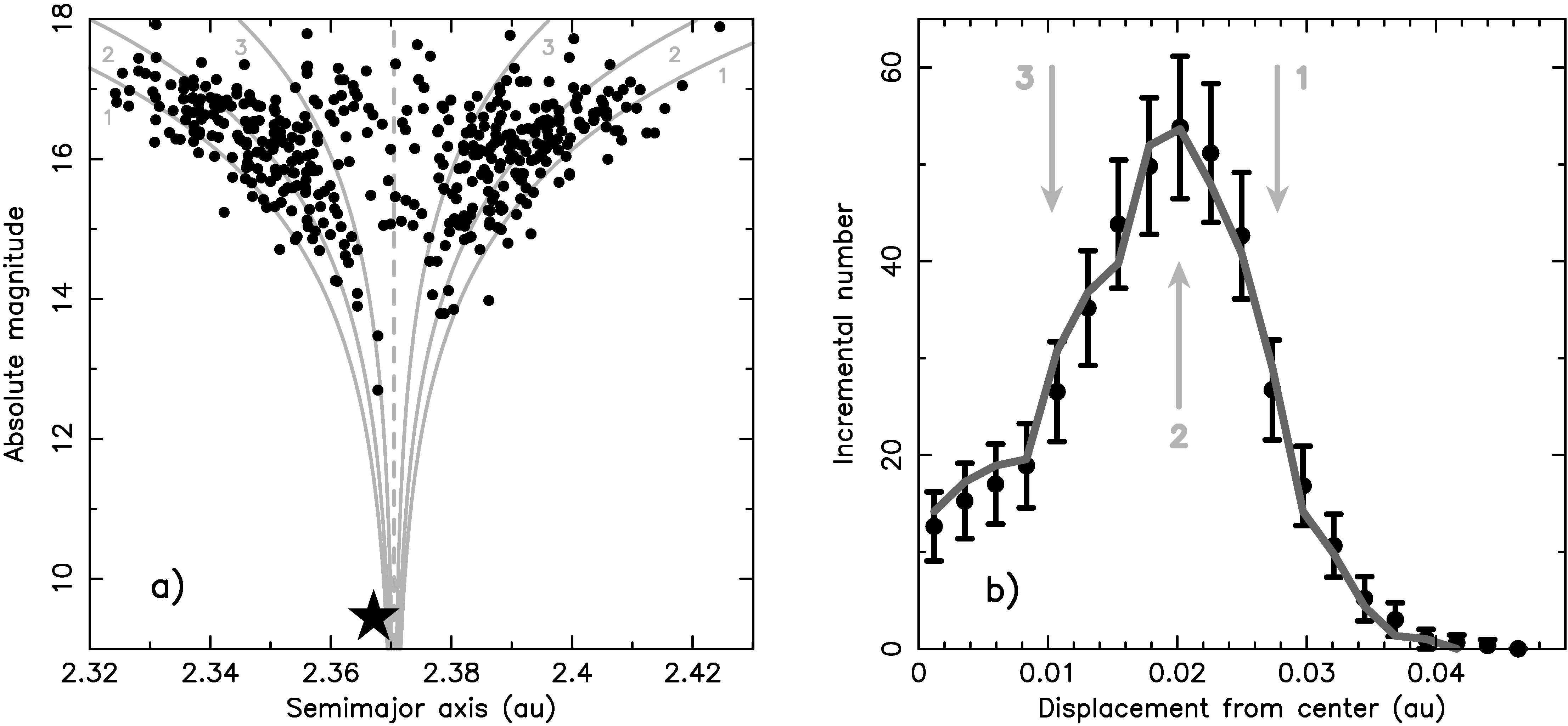}
 \caption{\small Left: The Erigone family members projected on the plane
  of the proper semimajor axis $a_{\rm P}$ and the absolute magnitude $H$; 432 numbered
  family members, including (163) Erigone (star), shown as black symbols.
  The gray lines show $0.2\,H = {\rm log}(|a_{\rm P}-a_0|/C)$, with $a_0=2.3705$~au
  and three different values of the $C$ parameter labeled 1, 2 and 3. Right: Fixing the
  $H$ level (16~magnitude in our case), one has a one-to-one link between
  the $C$ value and a displacement from the center $a_0$, shown here at the
  abscissa. The symbols represent the Erigone family using a statistical
  distribution in the $C$-bins (assuming a symmetry $C\rightarrow -C$ in
  this case); uncertainty is simply $\sqrt{N}$, where $N$ is the number of
  asteroids in the bin. A numerical model (dark gray line) seeks to match
  the distribution by adjusting several free parameters such as the family
  age and initial dispersal of fragments from the largest fragment. The
  gray arrows point to the corresponding $C=$~const. lines on the left panel.
  Adapted from {\em Vokrouhlick\'y et~al.} (2006a), 
  with the family update as of April 2014.}  
 \label{feri}
\end{figure*}

While the methods of dating young asteroid families involve
convergence of the orbital angles only, the determination of
ages of the asteroid pairs (e.g., {\em Vokrouhlick\'y and Nesvorn\'y},
2008; {\em Pravec et~al.}, 2010) represents an even more ambitious task. 
In this case, one seeks
to achieve a {\em full convergence} of two asteroidal orbits into a single
location in the Cartesian space (within the distance of about a radius of the
Hill sphere of the parent body) and with a small relative velocity
(comparable to the escape velocity from the parent body). It is not
surprising that the Yarkovsky effect again plays important role
in this effort. The best cases, such as the pair (6070)~Rheinland and 
(54827) 2001~NQ8, allow one to also infer constraints on the obliquities
of the individual components, consequently providing predictions directly 
testable by further observations (e.g., {\em Vokrouhlick\'y and Nesvorn\'y},
2009; {\em Vokrouhlick\'y et~al.}, 2011).

\bigskip
\noindent
 \textbf{4.5 Spreading of asteroid families}
\bigskip

Older asteroid families (ages $>10$~m.y., say) do not permit application
of the fine age-determination methods described in Sec.~4.4.
This is because orbits in the main asteroid belt are affected by 
deterministic chaos over long timescales. Hence it is not
possible to reliably reconstruct past values of the orbital secular 
angles, with the proper values of semimajor axis $a_{\rm P}$, eccentricity $e_{\rm P}$
and inclination $i_{\rm P}$ being the only well-defined parameters at hand. Still,
these proper elements are constructed using approximate dynamical models,
spanning time intervals quite shorter than the typical ages of large
asteroid families. While the deterministic chaos is still in action
over long timescales and produces a slow diffusion of the proper $e_{\rm P}$ and
$i_{\rm P}$ values, the Yarkovsky effect is the principal phenomenon that 
changes the proper $a_{\rm P}$ values of multi-kilometer-size asteroids.
{\em Bottke et~al.} (2001), studying an anomalous structure of the
Koronis family, presented the first clear example of the Yarkovsky effect 
sculpting a large-scale shape of an asteroid family in $a_{\rm P}$ and 
$e_{\rm P}$. It also approximately
constrained its age to $\sim (2.5-3)$~b.y. (see also {\em Vokrouhlick\'y et~al.}
(2010) for a similar study of the Sylvia family).

A novel method suitable for age-determination of families a few hundred
m.y. old has been presented by {\em Vokrouhlick\'y et~al.} (2006a). It
stems from the observation that small asteroids in some families are
pushed towards extreme values of the semimajor axis and, if plotted in
the $a_{\rm P}$ vs $H$ (absolute magnitude) diagram, they acquire an ``eared''
structure (Fig.~\ref{feri}). Since this peculiar structure is not 
compatible with a direct emplacement by any reasonable ejection field,
{\em Vokrouhlick\'y et~al.} (2006a) argued it must result from a 
long-term dynamical evolution of the family. In particular, postulating
that the initial dispersal in $a_{\rm P}$ of the family members was actually small,
they showed that Yarkovsky drift itself accounted for most of the family's 
extension in semimajor axis. Assisted by the YORP effect, which over a YORP-cycle
timescale tilts obliquities toward extreme values, the Yarkovsky
effect (dominated by its diurnal component) is maximized, and pushes
small family members towards the extreme values in $a_{\rm P}$. If properly
modeled, this method allows to approximately constrain the interval of time needed
since the family-forming event to reach the observed extension (Fig.~\ref{feri}).
Several applications of this method can be found in {\em Vokrouhlick\'y
et~al.} (2006a,b,c), {\em Bottke et~al.} (2007), {\em Carruba} (2009) or
{\em Carruba and Morbidelli} (2011). Recently {\em Bottke et~al.} (2015)
noticed that the classical setting of this method does not permit
a satisfactory solution for the low-albedo, inner-belt Eulalia family.
Their proposed modification requires an extended time spent by
small asteroids in the extreme obliquity state, which in turn requires
a simultaneous slow-down in the evolution of their rotation rates by
the YORP effect. In fact, this may be readily obtained by postulating
that the YORP strength changes on a timescale shorter than the YORP
cycle, an assumption that may follow from the extreme sensitivity
of the YORP effect to asteroid shape (the self-limitation effect discussed in
Sec.~2.1, also {\em Cotto-Figueroa et~al.}, 2015). It is not clear, however, 
why this phenomenon should manifest itself primarily in this particular 
family, or whether it generally concerns all families with $\sim$~b.y. age.

The model of {\em Vokrouhlick\'y et~al.} inherently contains a prediction 
that the small members in
the ``eared'' families have preferred obliquity values (such that
prograde-rotating objects occupy regions in the family with largest
$a$ values, and vice versa). Interestingly, recent works of {\em Hanu\v{s} et~al.}
(2013b) and {\em Kryszczy\'{n}ska} (2013) confirm this trend in the cases 
of several families, and more detailed studies are under way.

A peculiar situation arises for families embedded in the first-order
mean motion resonances with Jupiter. In these cases, the resonant lock 
prohibits large changes in the semimajor axis, but the Yarkovsky effect manifests
itself by a secular increase or decrease of the eccentricity. 
Modeling of this evolution allowed {\em Bro\v{z} and Vokrouhlick\'y}
(2008) and {\em Bro\v{z} et~al.} (2011) to estimate the age of the Schubart 
and Hilda families located in the 3/2 mean motion resonance with Jupiter.
\bigskip

\begin{center}
 \textbf{5. APPLICATIONS OF THE YORP AND BYORP EFFECTS}
\end{center}
\begin{figure}[t]
 \epsscale{1.}
 \plotone{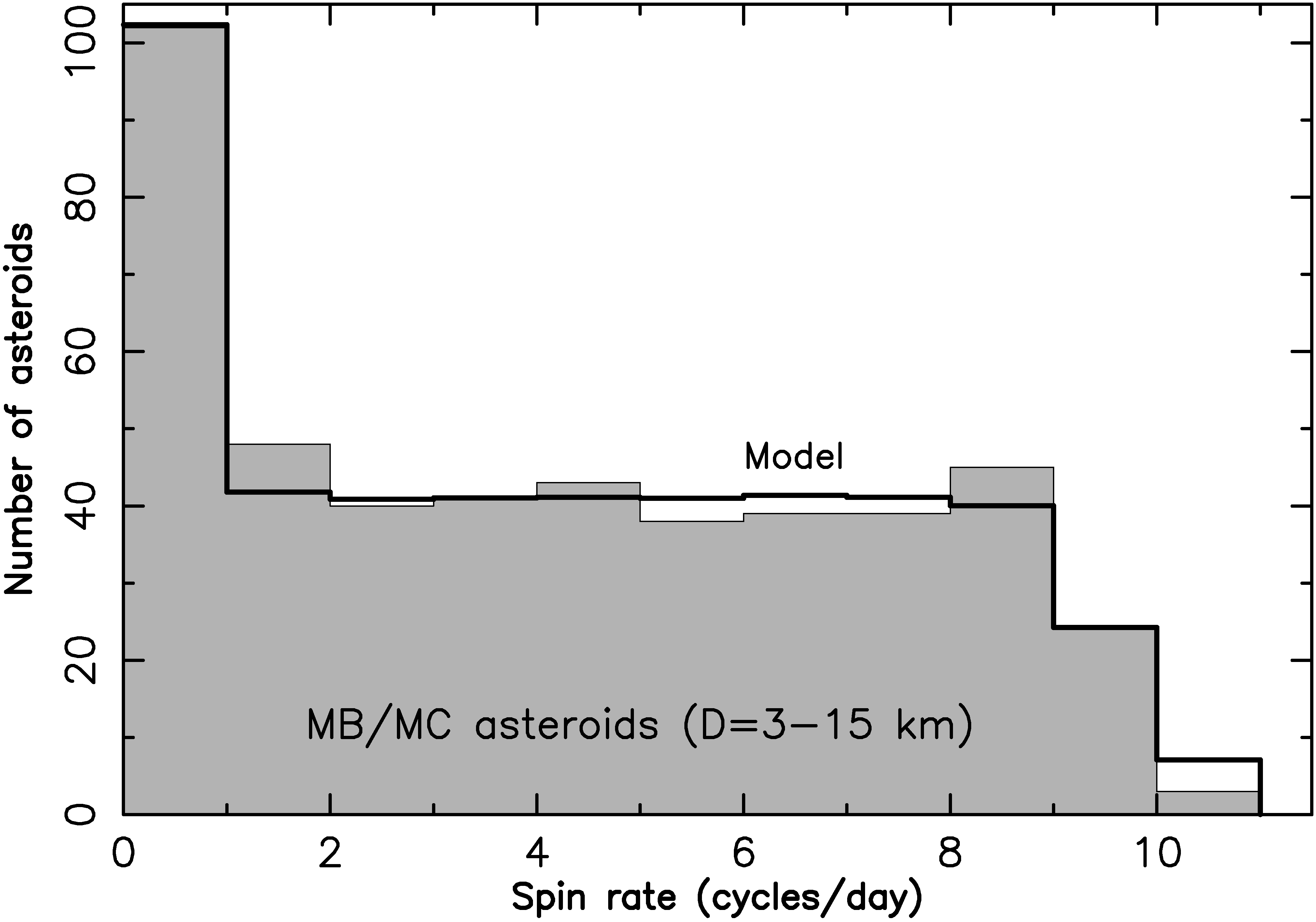}
 \caption{\small Spin rate distribution of 462 small main-belt and
  Mars-crossing asteroids (sizes in the 3-15~km range, with a median
  value of 6.5~km). The distribution is flat with only two features:
  (i) an excess of slow rotators with periods longer than 1~day (the
  first bin), and (ii) linear decrease on the 8 to 10 cycles/day interval.
  The latter is simply due to rotational fission limit dependence on
  the actual shape of the body, while the former holds information how the
  spin re-emerges from the slow-rotation limit. Results from a simple
  model of a YORP-relaxed population of objects is shown in black (model).
  Adapted from {\em Pravec et~al.} (2008), with an update from Petr
  Pravec as of April 2014.}  
 \label{fo}
\end{figure}
\begin{figure}[t]
 \epsscale{1.}
 \plotone{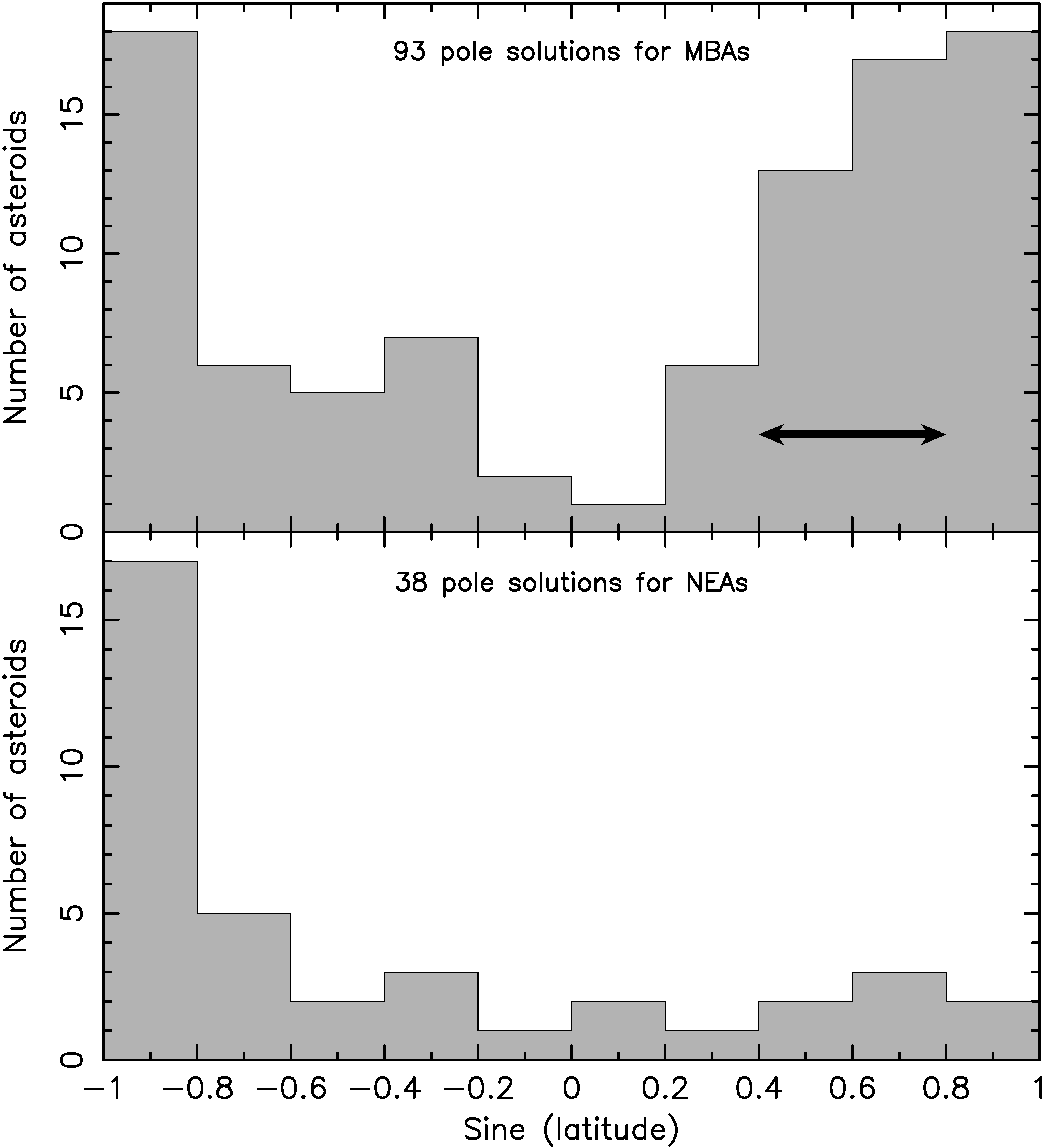}
 \caption{\small Top: Distribution of ecliptic pole latitude for 93 small
  main belt asteroids (MBAs; sizes less then $30$~km). The arrow indicates the
  zone of prograde-rotating objects potentially affected by the spin-orbit
  resonances (e.g., {\em Vokrouhlick\'y et~al.}, 2006d). This effect is
  nonexistent for retrograde-rotating objects and the poles are let to drift
  closer to the extreme value.
  Bottom: Distribution of ecliptic pole latitude for 38 near-Earth asteroids
  (NEAs). This is dominated by retrograde-rotating objects ($\simeq 73$\%
  cases), because this sense of rotation offers better chance to migrate
  to the planet-crossing space. In both cases, the tendency to extreme latitude
  values is due to the YORP effect. MBA data adapted from {\em Hanu\v{s} et~al.}
  (2013a), NEA data from the LCDB compilation as of February~2014.}  
 \label{fg}
\end{figure}

\bigskip
\noindent
 \textbf{5.1 Distribution of rotation rate and obliquity for
  small asteroids}
\bigskip

As explained in Sec.~2.1, a secular change in rotation rate and
obliquity are the two main dynamical implications of the YORP
effect. Therefore, it is has been natural to seek traits of
these trends among the populations of small asteroids. Luckily, the
amount of data and their quality have significantly increased over
the last decade and allowed such analyses.
\smallskip

\noindent{\bf Rotation-rate distribution.-- }The 
distribution of rotation frequencies of large asteroids in the main belt
matches a Maxwellian function quite well with a mean rotation period of
$\sim (8-12)$~hr, depending on the size of the bin used. However, data
for asteroids smaller than $\sim 20$~km show significant deviations from
this law, with many asteroids either having very slow or very fast
rotation rates. Note that similar data are also available for near-Earth
asteroids, but the main belt sample is more suitable because
its interpretation is not complicated by possible effects of
planetary close approaches. After eliminating known or suspected binary systems, 
solitary kilometer-size asteroids in the main asteroid belt were shown to 
have a roughly uniform distribution of rotation frequencies ({\em Pravec
et~al.}, 2008, and Fig.~\ref{fo}). The only statistically significant
deviation was an excess of slow rotators (periods less than a day or so).
Note that the sample described by {\em Pravec et~al.} (2008) is
superior to other existing datasets so far in elimination of all
possible survey biases (which may prevent recognition of slow rotators; 
P.~Pravec, private communication).

These results are well explained with a simple model of a relaxed YORP
evolution. In this view asteroid spin rates are driven by the YORP effect
toward extreme (large or small) values on a characteristic (YORP-)
timescale dependent on the size. Asteroids evolving toward a state of
rapid rotation shed mass and thus put a brake on their rotation rate, while those
who slow their rotation too much enter into a tumbling phase. They may
later emerge from this state naturally, with a new spin vector, or may gain
rotation angular momentum by sub-catastrophic impacts. After a few cycles
the spin rates settle to an approximately uniform distribution and the
memory of its initial value is erased. In fact, the observations similar
to those shown in Fig.~\ref{fo} may help to quantitatively calibrate the
processes that allow bodies to re-emerge from the slow-rotating state.

{\em Statler et~al.} (2013) presented a first attempt to obtain unbiased
rotation properties of very small near-Earth asteroids. They found an
anomalously large fraction of very fast rotating bodies in the group
having size $< 60$~m, which may witness a preferential ability of YORP
to accelerate small-asteroids' rotation rate. A larger sample, less vulnerable
to potential errors and biases, will be needed to verify this potentially
important result.
\smallskip

\noindent{\bf Obliquity distribution.-- }Similarly, the 
distribution of pole orientation of large asteroids in the
main belt is roughly isotropic, with only a moderate excess of prograde
rotating bodies. On the other hand, rotation poles of small asteroids
(sizes $\leq 30$~km) are strongly concentrated toward ecliptic south and
north poles ({\em Hanu\v{s} et~al.}, 2013a, and Fig.~\ref{fg}). Note
that this trend is better exhibited in the retrograde-rotating group
(obliquities $> 90^\circ$), because the prograde-rotating asteroids are
perturbed by secular spin-orbit resonances (e.g., {\em Vokrouhlick\'y 
et~al.}, 2006d). As a result, there is more mixing among the obliquities
$< 90^\circ$ which causes their flatter distribution in Fig.~\ref{fg}.
Overall, this result can again be matched with the above-mentioned
simple model of YORP evolution, because YORP torques drive obliquity
toward its extreme values (e.g., {\em \v{C}apek and Vokrouhlick\'y}, 2004).

The pole distribution of near-Earth asteroids, in
spite of a still limited sample, indicates a strong preference of directions
near the south ecliptic pole ({\em La Spina et~al.}, 2004, and Fig.~\ref{fg}). 
The ratio between the number of retrograde vs prograde rotating bodies is 
nearly 3:1. This is in a very good agreement with prediction from
a model, where most of near-Earth asteroids are delivered from the
main belt via principal resonant routes, secular $\nu_6$ resonance and 
3/1 mean motion resonance with Jupiter ({\em Bottke et~al.}, 2002b),
resupplied by the Yarkovsky effect ({\em Morbidelli and Vokrouhlick\'y},
2003). This is because while the 3/1 resonance may be reached from
heliocentric orbits with both larger and smaller value of the semimajor
axis, asteroids can enter the $\nu_6$ resonance only by decreasing their
semimajor axis. Taking into account the proportion by which these resonances
contribute to the near-Earth asteroid population ({\em Bottke et~al.}, 2002b),
one obtains the observed 3:1 ratio between spin retrograde vs prograde
rotators. This obviously assumes, the rotation pole directions do not 
become significantly modified after the asteroids enter the planet-crossing
zone.

Another interesting piece of information comes from a study of orbital pole 
distribution of small binary systems in the main belt. {\em Pravec et~al.} 
(2012) show that poles of these systems are non-isotropic with
strong concentration toward the ecliptic poles, mimicking thus the spin
distribution of solitary asteroids in the same class. This
picture is consistent with a model in which these small binaries are
formed by fission of the parent body, whose rotation has been brought
to the rotational limit by the YORP effect.

\bigskip
\noindent
 \textbf{5.2 Asteroids with rotation axes caught in spin-orbit resonances}
\bigskip

In an attempt to generalize Cassini's second and third laws, Giuseppe Colombo 
developed a mathematical model in the 1960's that describes the evolution of a
body's spin axis rotating about a principal axis of its inertia tensor
({\em Colombo}, 1966). Colombo included two fundamental
elements in his approach: (i) gravitational torques due to a massive center
(e.g., Sun), and (ii) regular precession of the orbital plane of the body
by exterior perturbers (e.g., planets). Because (i) produces a regular
precession of the spin axis, a secular spin-orbit resonance (with a stable
fixed point called Cassini state~2) may occur between its frequency
and the frequency by which
the orbital plane precesses in the inertial space. Such a resonance may
occur only for a certain range of obliquity and rotation period values,
and thus there is only a small probability that the spin state of any
given asteroid is located in the Cassini state~2 associated with one
of the frequencies by which its orbital plane precesses in space.

With this as background, the discovery of five prograde-rotating
Koronis member asteroids with similar spin vectors (i.e., spin axes
nearly parallel in inertial space and similar
rotation periods) was a surprise. Additionally, the sample of
retrograde-rotating asteroids in the same observation campaign showed
obliquities anomalously large ($\geq 154^\circ$) and either short or long
rotation periods ({\em Slivan}, 2002; {\em Slivan et~al.}, 2009). This 
puzzling situation, however, was
solved with a model where the gravitational spin dynamics was
complemented with the long-term effects of YORP torques ({\em Vokrouhlick\'y
et~al.}, 2003). The YORP effect was shown to bring, on a $\sim (2-3)$~Gy
timescale, prograde states close to Cassini state~2 associated with the
prominent $s_6$ frequency in the orbital precession, providing thus
a natural explanation for the alignment in inertial space. Note that
while the capture is fundamentally unstable, the evolution becomes
slowed down near the observed obliquities where the YORP effect changes
rotation period only slowly (Sec.~2.1). No resonant trapping zone exists for 
retrograde-rotating bodies, whose evolution is thus simpler and,
driven solely by the YORP effect, evolve toward extreme values 
in both their obliquities and rotation periods.

The possibility exists for asteroid spin states to be trapped in
similar spin-orbit resonant states, dubbed ``Slivan states'', for bodies 
residing on low-inclination orbits, especially in the central and 
outer parts of the main asteroid belt. Recently reported Slivan states 
in the inner part of the belt, namely in the Flora region ({\em
Kryszczy\'{n}ska}, 2013), are questionable because of their
instability. Yet, model refinment would be clearly needed
if more bodies are observed near these states in the Flora region.

\bigskip
\noindent
 \textbf{5.3 Formation and long-term evolution of binary systems}
\bigskip

The BYORP effect is predicted to play a fundamental role in the evolution 
of asteroid binaries. As noted earlier, it has been hypothesized that nearly 
all observed small, rubble-pile binary asteroid systems lie in an equilibrium 
state where BYORP and tidal torques are balanced. 

The BYORP effect plays many other roles in controlling the evolution of 
a binary asteroid. {\em Jacobson and Scheeres} (2011a) studied the 
evolution of asteroid systems arising from the rotational fission of a 
primary body (due  to YORP torques). While the initial creation of a stable 
binary  system is a complex process ({\em Walsh et al.}, this volume), once a stable 
binary forms with at least one of the bodies being synchronous, the BYORP 
effect can take control of its subsequent evolution. There are several 
different pathways, which we briefly review here. 

First, if the ratio between the 
secondary and primary is greater than $\sim 0.2$, the system is expected to 
eventually settle into a double-synchronous binary asteroid such as 
(69230)~Hermes. 
In this configuration both of the synchronous bodies can contribute to the 
BYORP effect, either working together to contract or expand the system, 
or working against each other. In none of these cases is it expected that 
the system will settle into a stable equilibrium, as only if the two BYORP 
effects counteract each other exactly would migration stop. Similarly, there 
are no significant tidal dissipation effects once a system is doubly-synchronous, 
and thus the case of contraction will lead directly to collapse (e.g., {\em Taylor
and Margot}, 2014). The expansion phase of a doubly-synchronous binary 
asteroid has not been investigated in detail as of yet in terms of physical 
evolution. However, as the system becomes larger it should be more 
susceptible to other exogenous perturbations (e.g., {\em Fang and Margot}, 2012). 

For stable binaries that have a mass ratio $< 0.2$ between the secondary 
and primary, the evolutionary path is seen to be quite different (see also
{\em Walsh et al.}, this volume).
If a stable binary is formed, it is generally a singly-synchronous system with 
the secondary in a synchronous state and the primary rotating faster than 
the spin rate. If the secondary's BYORP coefficient is negative and the 
system contracts, then it should migrate into a BYORP-tide equilibrium. Once 
in this state it may persist for long periods of time, as the system has been 
hardened against exogenous perturbations due to its more compact state 
(e.g., {\em Fang and Margot}, 2012). A noticeable outcome is that the primary 
body should lose spin rate, due to the tidal transfer of torque. However, the 
primary may still be subject to the YORP effect and thus may not exhibit a clear 
slowing of its spin rate. 

If the secondary's BYORP coefficient is positive the system expands, with 
tides now working in the same direction. In this case there is also an 
interesting interplay between the libration of the secondary about its 
synchronous state and tidal dissipation that acts to damp out such 
librations. In the expansive case without librational damping, the 
amplitude of libration is expected to increase as the orbit increases, due 
to an adiabatic integral involving the libration state (e.g., {\em Jacobson et~al.},
2014). How these two effects combine can control when the secondary can 
lose synchronous lock, causing the BYORP effect to shut down. The model and 
simulations developed in {\em Jacobson et al.} (2014) indicate that 
synchronicity is lost at a far enough distance so that further tidal 
evolution of the system does not occur, and the system can be described 
as a wide-asynchronous binary. This paper makes favorable comparisons 
between predictions of the theory and such observed binary systems. An 
alternate, earlier theory was proposed by {\em \'Cuk and Nesvorn\'y} (2010) 
in which the expanding system can become trapped in a resonance with the 
eccentricity of the orbit growing secularly. They hypothesized that such a 
system would then lose synchronicity, but subsequently relax back into 
synchronous rotation several times until the system enters a contractive 
phase. The very different predictions from these models indicate that 
the full interaction of such expanding binary systems is not yet fully 
understood. 

In addition to expansion and contraction effects, there may also be out 
of plane BYORP effects that cause migration of the binary system's 
orbit pole, similar to the YORP effect (see already {\em \'Cuk and Burns}, 
2005). {\em Steinberg and Sari} (2011) further studied these situations 
and proposed that, similar to YORP, the obliquity states should preferentially 
migrate towards some asymptotic values (either $0^\circ$, $90^\circ$ or 
$180^\circ$ in their model). {\em \'Cuk} (2007) 
noted that this effect, when combined with the characteristic zero-crossing 
of the BYORP coefficient as a function of obliquity, could create an 
accumulation of binaries at obliquities between these limits. It should 
also be mentioned that {\em McMahon and Scheeres} (2010b) did not predict 
an inclination evolution due to the BYORP effect, due to the effects of the 
primary oblateness. Thus, it is apparent that the obliquity migration of a 
binary orbit due to BYORP is not fully understood or settled, and remains a 
ripe topic for further investigation.  
\bigskip

\centerline{\textbf{6. CONCLUSION AND FUTURE WORK}}
\bigskip

As it is with many mature disciplines in science, studies of the Yarkovsky 
and YORP effects have their own agenda of development in the future years.
What makes them even more appealing is that some of these future
results have interesting implications for other domains in planetary
astronomy. Here we try to summarize at least a few examples.

While it seems nearly certain that numerous detections of the Yarkovsky 
effect will emerge from current and upcoming astrometric surveys
in the next decade (e.g., {\em Delb\`o et~al.}, 2008; {\em Mouret and Mignard}, 
2011; {\em Nugent et~al.}, 2012b; {\em Desmars}, 2015), more work is needed 
to secure YORP detections, especially across the whole range of possible 
rotation periods. This should help us understand in what proportion the 
YORP effect results in acceleration or deceleration of the rotation rate.

The binary YORP detections are in their infancy but will become an important topic
of future research. This is because the BYORP effect is an essential
element, as far as we understand it today, in orbital evolution of binary 
asteroids. Detections, or continuing non-detections, of the expected 
BYORP signal will have implications not only for the orbital evolution 
pathways of binaries and their physical parameters, but also for estimates 
of their lifetime and formation rate.

As asteroids' rotations become slower by the YORP effect, they
naturally enter the tumbling state. The available models so far,
whether analytical or numerical in nature ({\em Vokrouhlick\'y et~al.},
2007; {\em Cical\`o and Scheeres}, 2010; {\em Breiter et~al.}, 2011),
indicate that the YORP effect keeps navigating the rotation through
the tumbling phase space without an easy return to the rotation
about the shortest axis of the inertia tensor. Yet, more than 90\%
of asteroids do rotate in the shortest-axis mode. Solution of this
conundrum is not clear yet and warrants further work. The above
mentioned models of the YORP effect in the tumbling regime neglect thermal
inertia, which may be an important factor. Additionally, no detailed
model combining the YORP effect and the effects of inelastic energy
dissipation inside the body has been presented (though the initial work does
not seem to remedy the problem; S.~Breiter, private communication).

While it is generally accepted that the YORP effect is the driving dynamical 
process that brings small asteroids to their fission, more work is needed
to understand how the fission mechanics really works. Along the path to the fission
limit, the body may undergo structural and shape changes that could
either help the fission process, or potentially invert the YORP 
acceleration effectively prevent fission. 
It is not known which of these alternatives typically 
dominates and in what proportion. This, again, could have important 
implications for the formation rate of both small binaries and asteroid pairs.
Additionally, this would help better understand the YORP self-limitation
processes and the way how they potentially modify classical YORP results.

\medskip
\noindent
\textbf{Acknowledgments.} The authors are grateful to Davide Farnocchia
 for assistance in the compilation of Table~1 and Petr Pravec for providing
 data shown in Figure~8. We also thank reviewers, David P. Rubincam and Ben
 Rozitis, for their suggestions that helped to improve the original form of
 the chapter. The work of DV was partially
 supported by the Czech Grant Agency (grant P209-13-01308S).
 WFB participation was supported by NASA's Solar System Evolution Research
 Virtual Institute (SSERVI) program through a grant to the Institute for
 the Science of Exploration Targets at the Southwest Research Institute in
 Boulder, CO. The work of SRC was conducted at the Jet Propulsion Laboratory,
 California Institute of Technology, under a contract with NASA. TSS acknowledges
 support from NASA Planetary Geology \& Geophysics grant NNX11AP15G.

\bigskip

\centerline\textbf{ REFERENCES}
\bigskip
\parskip=0pt
{\small
\baselineskip=11pt

\refs Afonso, G. B., Gomes, R. S., and Florczak, M. A. (1995) Asteroid fragments
 in Earth-crossing orbits. {\em Planet. Sp. Sci., 43}, 787-795.

\refs Beekman, G. (2006) I.O. Yarkovsky and the discovery of 'his' effect.
 {\em J. Hist. Astron., 37}, 71-86.

\refs Bottke, W. F., Vokrouhlick\'y, D., and Nesvorn\'y, D. (2007)
 An asteroid breakup 160 My ago as a probable source of the K-T impactor.
 {\em Nature, 449}, 48-53.

\refs Bottke, W. F., Vokrouhlick\'y, D., Bro\v{z}, M., Nesvorn\'y, D., and
 Morbidelli A. (2001) Dynamical spreading of asteroid families via
 the Yarkovsky effect: The Koronis family and beyond. {\em Science, 294},
 1693-1696.

\refs Bottke, W. F., Vokrouhlick\'y, D., Rubincam, D. P., Bro\v{z}, M. (2002a)
 Dynamical evolution of asteroids and meteoroids using the Yarkovsky effect.
 In {\em Asteroids III} (W. F. Bottke et~al., eds.), Univ. of Arizona Press,
 Tucson, 395-408. 

\refs Bottke, W. F., Vokrouhlick\'y, D., Rubincam, D. P., and Nesvorn\'y, D.
 (2006) The Yarkovsky and Yorp effects: Implications for asteroid
 dynamics. {\em Annu. Rev. Earth Planet. Sci., 34}, 157-191.

\refs Bottke, W. F., Morbidelli, A., Jedicke, R., et~al. (2002b) Debiased
 orbital and absolute magnitude distribution of the near-Earth objects.
 {\em Icarus, 156}, 399-433.

\refs Bottke, W. F., Vokrouhlick\'y, D., Walsh, K., et~al. (2015) In search
 of the source of asteroid (101955) Bennu: Application of the stochastic YORP
 model. {\em Icarus, 247}, 191-217.

\refs Breiter, S., and Micha{\l}ska, H. (2008) YORP torque as the function
 of shape harmonics. {\em Mon. Not. R. Astron. Soc., 388}, 927-944.

\refs Breiter, S., and Vokrouhlick\'y, D. (2011) YORP effect with
 anisotropic radiation. {\em Mon. Not. R. Astron. Soc., 410},
 2807-2816.

\refs Breiter, S., Vokrouhlick\'y, D., and Nesvorn\'y D. (2010a) Analytical
 YORP torques model with an improved temperature distribution function.
 {\em Mon. Not. R. Astron. Soc., 401}, 1933-1949.

\refs Breiter, S., Bartczak, P., and Czekaj, M. (2010b) YORP torques with
 1D thermal model. {\em Mon. Not. R. Astron. Soc., 408}, 1576-1589.

\refs Breiter, S., Ro\.zek, A., and Vokrouhlick\'y, D. (2011) YORP effect
 on tumbling objects. {\em Mon. Not. R. Astron. Soc., 417}, 2478-2499.

\refs Breiter, S., Micha{\l}ska, H., Vokrouhlick\'y, D., and Borczyk, W. (2007)
 Radiation induced torques on spheroids. {\em Astron. Astrophys., 471},
 345-353.

\refs Breiter, S., Bartczak, P., Czekaj, M., Oczujda, B., and
 Vokrouhlick\'y, D. (2009) The YORP effect on 25143 Itokawa. {\em
 Astron. Astrophys., 507}, 1073-1081.

\refs Bro\v{z}, M., and Vokrouhlick\'y, D. (2008), Asteroids in the first
 order resonances with Jupiter. {\em Mon. Not. R. Astron. Soc., 390},
 715-732.

\refs Bro\v{z}, M., Vokrouhlick\'y, D., Morbidelli, A., Nesvorn\'y, D.,
 and Bottke, W. F. (2011), Did the Hilda collisional family form during the
 late heavy bombardment? {\em Mon. Not. R. Astron. Soc., 414}, 2716-2727.

\refs \v{C}apek, D., and Vokrouhlick\'y, D. (2004) The YORP effect with
 finite thermal conductivity. {\em Icarus, 172}, 526-536.

\refs Carruba, V. (2009) The (not so) peculiar case of the Padua family.
 {\em Mon. Not. R. Astron. Soc., 395}, 358-377.

\refs Carruba, V., and Morbidelli, A. (2011) On the first $\nu_6$ anti-aligned
 librating asteroid family of Tina. {\em Mon. Not. R. Astron. Soc., 412},
 2040-2051.

\refs Chesley, S. R., Ostro, S. J., Vokrouhlick\'y, D., et~al. (2003)
 Direct detection of the Yarkovsky effect via radar ranging to
 the near-Earth asteroid 6489 Golevka. {\em Science, 302},
 1739-1742.

\refs Chesley, S. R., Vokrouhlick\'y, D., Ostro, S. J., et~al. (2008)
 Direct estimation of Yarkovsky accelerations on near-Earth asteroids.
 In {\em Asteroids, Comets, Meteors}, paper id. 8330
 (http://\lb{2}adsabs.\lb{2}harvard.\lb{2}edu/\lb{2}abs/\lb{2}2008LPICo1405.8330C).

\refs Chesley, S. R., Farnocchia, D., Nolan, M. C., et~al. (2014) Orbit
 and bulk density of the OSIRIS-REx target asteroid (101955)
 Bennu. {\em Icarus, 235} 5-22.

\refs Cical\`{o}, S., and Scheeres, D. J. (2010) Averaged rotational dynamics
 of an asteroid in tumbling rotation under the YORP torque. {\em Celest.
 Mech. Dyn. Astron., 106}, 301-337.

\refs Colombo, G. (1966) Cassini's second and third laws. {\em Astron. J.,
 71}, 891-896.

\refs Cotto-Figueroa, D. Statler, T. S., Richardson, D. C., and Tanga, P.
 (2015) Coupled spin and shape evolution of small rubble-pile asteroids: 
 self-limitation of the YORP effect. {\em Astrophys. J.}, in press. 

\refs \'{C}uk, M. (2007) Formation and destruction of small binary asteroids.
 {\em Astrophys. J. Lett., 659}, L57-L60.

\refs \'{C}uk, M., and Burns J. A. (2005) Effects of thermal radiation on the
 dynamics of binary NEAs. {\em Icarus, 176}, 418-431.

\refs \'{C}uk, M., and Nesvorn\'y, D. (2010) Orbital evolution of
 small binary asteroids. {\em Icarus, 207}, 732-743.

\refs Delb\`o, M., Tanga, P., and Mignard, F. (2008) On the detection of the
 Yarkovsky effect on near-Earth asteroids by means of Gaia. {\em 
 Planet. Sp. Sci., 56}, 1823-1827.

\refs Desmars, J. (2015) Detection of Yarkovsky acceleration in the context
 of precovery observations and the future Gaia catalogue. {\em Astron.
 Astrophys.}, in press. 

\refs \v{D}urech, J. (2005) 433 Eros - comparison of lightcurve extrema
 from 1901-1931 with the present rotation state. {\em Astron. Astrophys., 
 431}, 381-383.

\refs \v{D}urech, J., Vokrouhlick\'y D., Kaasalainen M., et~al. (2008a)
 Detection of the YORP effect for asteroid (1620) Geographos.
 {\em Astron. Astrophys., 489}, L25-L28.

\refs \v{D}urech, J., Vokrouhlick\'y D., Kaasalainen M., et~al. (2008b)
 New photometric observations of asteroids (1862) Apollo and (25143) Itokawa -
 analysis of YORP effect. {\em Astron. Astrophys., 488}, 345-350.

\refs \v{D}urech, J., Vokrouhlick\'y D., Baransky A. R., et~al. (2012)
 Analysis of the rotation period of asteroids (1865) Cerberus,
 (2100) Ra-Shalom and (3103) Eger - search for the YORP effect.
 {\it Astron. Astrophys., 547}, A10(9pp)

\refs Fang, J., and Margot, J.-L. (2012) Near-earth binaries and
 triples: Origin and evolution of spin-orbital  properties.
 {\em Astron. J., 143}, 24(14pp).

\refs Farinella, P., Vokrouhlick\'y, D., and Hartmann, W. K. (1998)
 Meteorite delivery via Yarkovsky orbital drift. {\em Icarus, 132},
 378-387.

\refs Farnocchia, D., and Chesley, S. R. (2014) Assessment of the
 2880 impact threat from asteroid (29075) 1950 DA. {\em Icarus, 229},
 321-327. 

\refs Farnocchia, D., Chesley, S. R., Chodas, P. W., Micheli, M.,
 Tholen, D. J., Milani, A., Elliott, G. T., and Bernardi, F. (2013a)
 Yarkovsky-driven impact risk analysis for asteroid (99942) Apophis.
 {\em Icarus, 224}, 192-200.

\refs Farnocchia, D., Chesley, S. R., Vokrouhlick\'y, D., Milani, A.,
 Spoto, F., and Bottke, W. F. (2013b) Near Earth asteroids with
 measurable Yarkovsky effect. {\em Icarus, 224}, 1-13.

\refs Goldreich, P., and Sari, R. (2009) Tidal evolution of rubble
 piles. {\em Astrophys. J., 691}, 54-60.

\refs Golubov, O., and Krugly, Y. N. (2012) Tangential component of the YORP
 effect. {\em Astrophys. J. Lett., 752}, L11(5pp).

\refs Golubov, O., Scheeres, D. J., and Krugly, Y. N. (2014) A three-dimensional
 model of tangential YORP. {\em Astrophys. J., 794}, 22(9pp).

\refs Hanu\v{s}, J., \v{D}urech, J., Bro\v{z}, M., et~al. (2013a)
 Asteroids' physical models from combined dense and sparse photometry and
 scaling of the YORP effect by the observed obliquity distribution.
 {\em Astron. Astrophys., 551}, A67(16pp).

\refs Hanu\v{s}, J., Bro\v{z}, M., \v{D}urech, J., et~al. (2013b)
 An anisotropic distribution of spin vectors in asteroid families. {\em 
 Astron. Astrophys., 559}, A134(19pp).

\refs Hapke, B. (1993) Theory of reflectance and emittance spectroscopy.
 Cambridge University Press.

\refs Hudson, R. S., and Ostro, S. J. (1995) Shape and non-principal axis
 spin state of asteroid 4179 Toutatis. {\em Science, 270}, 84-86.

\refs Jacobson, S. A., and Scheeres, D. J. (2011a) Dynamics of
 rotationally fissioned asteroids: Source of observed small
 asteroid systems. {\em Icarus, 214}, 161-178.

\refs Jacobson, S. A., and Scheeres, D. J. (2011b) Long-term stable
 equilibria for synchronous binary asteroids. {\em Astrophys. J.
 Lett., 736}, L19(5pp).

\refs Jacobson, S. A., Scheeres, D. J., and McMahon, J. (2014) Formation
 of the wide asynchronous binary asteroid population. {\em Astrophys.
 J., 780}, 60(21pp).

\refs Kaasalainen, M., and Nortunen, H. (2013) Compact YORP formulation
 and stability analysis. {\em Astron. Astrophys., 558}, A104(8pp).

\refs Kaasalainen, M., \v{D}urech, J., Warner, B. D., Krugly, Y. N., and
 Gaftonyuk, N. M. (2007) Acceleration of the rotation of asteroid
 1862 Apollo by radiation torques. {\em Nature, 446}, 420-422.

\refs Kryszczy\'nska, A. (2013) Do Slivan states exist in the Flora
 family? II. Fingerprints of the Yarkovsky and YORP effects. 
 {\em Astron. Astrophys., 551}, A102(6pp).

\refs La Spina, A., Paolicchi, P., Kryszczy\'nska, A., and Pravec, P.
 (2004) Retrograde spins of near-Earth asteroids from the Yarkovsky effect.
 {\em Nature, 428}, 400-401.

\refs Lauretta, D. S., Barucci, M.A ., Binzel, R. P., et~al. (2015)
 The OSIRIS-REx target asteroid (101955) Bennu: Constraints on its physical,
 chemical, and dynamical nature from astronomical observations. {\em Meteor.
 Planet. Sci.}, in press.

\refs Lowry, S. C., Fitzsimmons, A., Pravec, P., et~al. (2007) Direct
 detection of the asteroidal YORP effect. {\em Science, 316}, 272-274.

\refs Lowry, S. C., Weissman, P. R., Duddy, S. R., et~al. (2014) 
 The internal structure of asteroid (25143) Itokawa as revealed by
 detection of YORP spin-up. {\em Astron. Astrophys., 562}, A48(9pp).

\refs McMahon, J., and Scheeres, D. J. (2010a) Secular orbit variation
 due to solar radiation effects: a detailed model for BYORP.
 {\em Cel. Mech. Dyn. Astron., 106}, 261-300.

\refs McMahon, J., and Scheeres, D. J. (2010b) Detailed prediction for
 the BYORP effect on binary near-Earth Asteroid (66391) 1999 KW4 and
 implications for the binary population. {\em Icarus, 209}, 494-509.

\refs Milani, A., Chesley, S. R., Sansaturio, M. E., Bernardi, F.,
 Valsecchi, G. B., and Arratia, O. (2009) Long term impact risk for (101955)
 1999~RQ36. {\em Icarus, 203}, 460-471.

\refs Mommert, M., Hora, J. L., Farnocchia, D., et~al. (2014) Constraining
 the physical properties of near-Earth object 2009~BD. {\em Astrophys.
 J., 786}, 148(9pp).

\refs Morbidelli, A., and Vokrouhlick\'y, D. (2003) The Yarkovsky-driven
 origin of near-Earth asteroids. {\em Icarus, 163},  120-134.

\refs Mouret, S., and Mignard, F. (2011) Detecting the Yarkovsky effect
 with the Gaia mission: list of the most promising candidates.
 {\em Mon. Not. R. Astron. Soc., 413}, 741-748.

\refs Nesvorn\'y, D., and Bottke, W. F. (2004) Detection of the
 Yarkovsky effect for main-belt asteroids. {\em Icarus, 170}, 324-342.

\refs Nesvorn\'y, D., and Vokrouhlick\'y, D. (2006) New candidates
 for recent asteroid breakups. {\em Astron. J., 132}, 1950-1958.

\refs Nesvorn\'y, D., and Vokrouhlick\'y, D. (2007) Analytic theory of
 the YORP effect for near-spherical objects. {\em Astron. J., 134},
 1750-1768.

\refs Nesvorn\'y, D., and Vokrouhlick\'y, D. (2008a) Analytic theory for
 the YORP effect on obliquity. {\em Astron. J., 136}, 291-299.

\refs Nesvorn\'y, D., and Vokrouhlick\'y, D. (2008b) Vanishing torque
 from radiation pressure. {\em Astron. Astrophys., 480}, 1-3.

\refs Nesvorn\'y, D., Vokrouhlick\'y, D., and Bottke, W. F. (2006)
 A main belt asteroid break-up 450 ky ago. {\em Science, 312}, 1490.

\refs Nesvorn\'y, D., Bottke, W. F., Dones, L., and Levison, H. F. (2002)
 The recent breakup of an asteroid in the main-belt region. {\em Nature,
 417}, 720-722,

\refs Nesvorn\'y, D., Bottke, W. F., Levison, H. F., and Dones, L. (2003)
 Recent origin of the zodiacal dust bands. {\em Astrophys. J., 591}, 486-497.

\refs Nesvorn\'y, D., Vokrouhlick\'y, D., Morbidelli, A., and Bottke, W. F.
 (2009) Asteroidal source of L chondrite meteorites. {\em Icarus,
 200}, 698-701.

\refs Nesvorn\'y, D., Bottke, W. F., Vokrouhlick\'y, D., Sykes, M., Lien, D. J.,
 and Stansberry, J. (2008) Origin of the near-ecliptic zodiacal dust band.
 {\em Astrophys. J. Lett., 679}, L143-L146.

\refs Novakovi\'{c}, B. (2010) Portrait of Theobalda as a young asteroid
 family. {\em Mon. Not. R. Astron. Soc., 407}, 1477-1486.

\refs Novakovi\'{c}, B., Dell'Oro, A., Cellino, A., and Kne\v{z}evi\'{c},
 Z. (2012) Recent collisional jet from a primitive asteroid.
 {\em Mon. Not. R. Astron. Soc., 425}, 338-346. 

\refs Novakovi\'{c}, B., Hsieh, H. H., Cellino, A., Micheli, M., and
 Pedani, M. (2014) Discovery of a young asteroid cluster associated with
 P/2012 F5 (Gibbs). {\em Icarus, 231}, 300-309.

\refs Nugent, C. R., Margot, J. L., Chesley, S. R., and Vokrouhlick\'y, D.
 (2012a) Detection of semi-major axis drifts in 54 near-Earth asteroids.
 {\em Astron. J., 144}, 60(13pp).

\refs Nugent, C. R., Mainzer, A., Masiero, J., Grav, T., and Bauer, J.
 (2012b) The Yarkovsky drift's influence on NEAs: Trends and predictions
 with NEOWISE measurements. {\em Astron. J., 144}, 75(6pp).

\refs \"Opik, E. J. (1951) Collision probabilities with the planets and
 the distribution of interplanetary matter. {\em Proc. Roy. Irish Acad.,
 A54}, 165-199.

\refs Ostro, S. J., Margot, J.-L., Benner, L. A. M., et~al. (2006)
 Radar imaging of binary near-Earth asteroid (66391) 1999~KW4. {\em
 Science, 314}, 1276-1280.

\refs Paddack, S. J. (1969) Rotational bursting of small celestial bodies:
 Effects of radiation pressure. {\em J. Geophys. Res., 74}, 4379-4381.

\refs Paddack, S. J., and Rhee, J. W. (1975) Rotational bursting of
 interplanetary dust particles. {\em Geophys. Res. Lett., 2}, 365-367.

\refs Peterson, C. (1976) A source mechanism for meteorites controlled
 by the Yarkovsky effect. {\em Icarus, 29}, 91-111.

\refs Pravec, P., Harris, A. W., Vokrouhlick\'y, D., et~al. (2008) Spin
 rate distribution of small asteroids. {\em Icarus, 197}, 497-504.

\refs Pravec, P., Vokrouhlick\'y, D., Polishook, D., et~al. (2010)
 Asteroid pairs formed by rotational fission. {\em Nature, 466},
 1085-1088.

\refs Pravec, P., Scheirich, P., Vokrouhlick\'y, D., et~al. (2012) Binary
 asteroid population. II. Anisotropic distribution of orbit poles. {\em
 Icarus, 218}, 125-143. 

\refs Pravec, P., Scheirich, P., \v{D}urech, J., et~al. (2014) The tumbling
 spin state of (99942) Apophis. {\em Icarus, 233}, 48-60. 

\refs Radzievskii, V. V. (1952a) A mechanism for the disintegration of
 asteroids and meteorites. {\em Astron. Zh., 29}, 162-170.

\refs Radzievskii, V. V. (1952b) The influence of anisotropy of
 re-emited sunlight on the orbital motion of asteroids and meteoroids.
 {\em Astron. Zh., 29}, 1952-1970.

\refs Richardson, D. C., Elankumaran, P., and Sanderson, R. E. (2005)
 Numerical experiments with rubble piles: equilibrium shapes and spins.
 {\em Icarus, 173}, 349-361.

\refs Rozitis, B., and Green, S. F. (2012) The influence of rough surface
 thermal-infrared beaming on the Yarkovsky and YORP effects.
 {\em Mon. Not. R. Astron. Soc., 423}, 367-388.

\refs Rozitis, B., and Green, S. F. (2013) The influence of global self-heating
 on the Yarkovsky and YORP effects. {\em Mon. Not. R. Astron. Soc., 433},
 603-621.

\refs Rozitis, B., and Green, S. F. (2014) Physical characterisation of near-Earth
 asteroid (1620) Geographos. Reconciling radar and thermal-infrared observations.
 {\em Astron. Astrophys., 568}, A43(11pp). 

\refs Rozitis, B., MacLennan, E., and Emery, J. P. (2014) Cohesive forces prevent
 the rotational breakup of rubble-pile asteroid (29075) 1950 DA. {\em Nature, 512},
 174-176.

\refs Rozitis, B., Duddy, S. R., Green, S. F., and Lowry, S. C. (2013) A
 thermophysical analysis of the (1862) Apollo Yarkovsky and YORP effects.
 {\em Astron. Astrophys., 555}, A20(12pp).

\refs Rubincam, D. P. (1982) On the secular decrease in the semimajor axis of
 LAGEOS's orbit. {\em Celest. Mech., 26}, 361-382.

\refs Rubincam, D. P. (1995) Asteroid orbit evolution due to thermal drag.
 {\em J. Geophys. Res., 100}, 1585-1594.

\refs Rubincam, D. P. (1998) Yarkovsky thermal drag on small asteroids
 and Mars-Earth delivery. {\em J. Geophys. Res., 103}, 1725-1732.

\refs Rubincam, D. P. (2000) Radiative spin-up and spin-down of small
 asteroids. {\em Icarus, 148}, 2-11.

\refs Rubincam, D. P., and Paddack, S. J. (2010) Zero secular torque on
 asteroids from impinging solar photons in the YORP effect: A simple
 proof. {\em Icarus, 209}, 863-865.

\refs Scheeres, D. J. (2007) The dynamical evolution of uniformly
 rotating asteroids subject to YORP. {\em Icarus, 188}, 430-450.

\refs Scheeres, D. J. (2015) Landslides and mass shedding on spinning
 spheroidal asteroids. {\em Icarus, 247}, 1-17.

\refs Scheeres, D. J., and Gaskell, R. W. (2008) Effect of density
 inhomogeneity on YORP: The case of Itokawa. {\em Icarus, 198}, 125-129. 

\refs Scheeres, D. J., and Mirrahimi, S. (2008) Rotational dynamics
 of a solar system body under solar radiation torques. {\em Celest. Mech.
 Dyn. Astron., 100}, 69-103.

\refs Scheeres, D. J., Hartzell, C. M.;, S\'anchez, P., and Swift, M.
 (2010) Scaling forces to asteroid surfaces: The role of cohesion.
 {\em Icarus, 210}, 968-984.

\refs Scheeres, D. J., Abe, M., Yoshikawa, M., Nakamura, R., Gaskell,
 R. W., and Abell, P. A. (2007) The effect of YORP on Itokawa. {\em
 Icarus, 188}, 425-429. 

\refs Scheirich, P., Pravec, P., Jacobson, S. A., et~al. (2015) The binary
 near-Earth asteroid (175706) 1996 FG3 - An observational constraint on
 its orbital stability. {\em Icarus, 245}, 56-63.

\refs Schwartz, S. R., Richardson, D. C., and Michel, P. (2012)
 An implementation of the soft-sphere discrete element method in a
 high-performance parallel gravity tree-code. {\em Granular Matter,
 14}, 363-380.

\refs \v{S}eve\v{c}ek, P., Bro\v{z}, M., \v{C}apek, D., and \v{D}urech, J.
 (2015) The thermal emission from boulders on (25143) Itokawa and 
 general implications for the YORP effect. {\em Mon. Not. R. Astron.
 Soc.}, submitted.

\refs Sekiya, M., and Shimoda, A. A. (2013) An iterative method for
 obtaining a nonlinear solution for the temperature distribution of
 a rotating spherical body revolving in a circular orbit around a star.
 {\em Planet. Sp. Sci., 84}, 112-121.

\refs Sekiya, M., and Shimoda, A. A. (2014) An iterative method for
 obtaining a nonlinear solution for the temperature distribution of
 a rotating spherical body revolving in an eccentric orbit.
 {\em Planet. Sp. Sci. 97}, 23-33.

\refs Slivan, S. M. (2002) Spin vector alignment of Koronis family
 asteroids. {\em Nature, 419}, 49-51.

\refs Slivan, S. M., Binzel, R. P., Kaasalainen, M. et~al. (2009) Spin
 vectors in the Koronis family. II. Additional clustered spins, and one
 stray. {\em Icarus, 200}, 514-530.

\refs Spitale, J., and Greenberg, R. (2001) Numerical evaluation of the
 general Yarkovsky effect: Effects on semimajor axis. {\em Icarus, 149},
 222-234.

\refs Spitale, J., and Greenberg, R. (2002) Numerical evaluation of the
 general Yarkovsky effect: Effects on eccentricity and longitude of periapse.
 {\em Icarus, 156}, 211-222.

\refs Statler, T. S. (2009) Extreme sensitivity of the YORP effect to
 small-scale topography. {\em Icarus, 202}, 502-513. 

\refs Statler, T. S., Cotto-Figueroa, D., Riethmiller, D. A., and Sweeney,
 K. M. (2013) Size matters: The rotation rates of small near-Earth asteroids.
 {\em Icarus, 225}, 141-155.

\refs Steinberg, E., and Sari, R. (2011) Binary YORP effect and evolution
 of binary asteroids. {\em Astron. J., 141}, 55(10pp).

\refs Taylor, P. A., and Margot, J. L. (2014) Tidal end states of binary
 asteroid systems with a nonspherical component. {\em Icarus, 229}, 418-422.

\refs Taylor, P. A., Margot, J. L., Vokrouhlick\'y, D., et~al. (2007) Spin
 rate of asteroid (54509) 2000 PH5 increasing due to the YORP
 effect. {\em Science, 316}, 274-277.

\refs Taylor, P. A., Nolan, M. C., Howell, E. S., et~al. (2012) 2004~FG11.
 {\em CBET, 3091}, 1.

\refs Vinogradova, V. P., and Radzievskii, V. V. (1965) The acceleration of
 the Martian satellites and the stabilization of orbits of artificial
 satellites. {\em Astron. Zh., 42}, 424-432.

\refs Vokrouhlick\'y, D. (1998a) Diurnal Yarkovsky effect for meter-sized
 asteroidal fragments' mobility I. Linear theory. {\em Astron. Astrophys.,
 335}, 1093-1100.

\refs Vokrouhlick\'y, D. (1998b) Diurnal Yarkovsky effect for meter-sized
 asteroidal fragments' mobility II. Non-sphericity effects. {\em Astron.
 Astrophys., 338}, 353-363.

\refs Vokrouhlick\'y, D. (1999) A complete linear model for the Yarkovsky
 thermal force on spherical asteroid fragments. {\em Astron. Astrophys.,
 344}, 362-366.

\refs Vokrouhlick\'y, D., and Bro\v{z}, M. (1999) An improved model of
 the seasonal Yarkovsky force for the regolith-covered asteroid fragments.
 {\em Astron. Astrophys., 350}, 1079-1084.

\refs Vokrouhlick\'y, D., and Farinella, P. (2000) Efficient delivery of
 meteorites to the Earth from a wide range of asteroid parent bodies.
 {\em Nature, 407}, 606-608.

\refs Vokrouhlick\'y, D., and Milani, A. (2000) Direct solar radiation
 pressure on the orbits of small near-Earth asteroids: observable effects?
 {\em Astron. Astrophys., 362}, 746–755.

\refs Vokrouhlick\'y, D., and Bottke, W. F. (2001) The Yarkovsky thermal
 force on small asteroids and their fragments: Choosing the right albedo.
 {\em Astron. Astrophys., 371}, 350-353.

\refs Vokrouhlick\'y, D., and \v{C}apek, D. (2002) YORP-induced long-term
 evolution of the spin state of small asteroids and meteoroids. I.
 Rubincam's approximation. {\em Icarus, 159}, 449-467.

\refs Vokrouhlick\'y, D., and Nesvorn\'y, D. (2008) Pairs of asteroids
 probably of common origin. {\em Astron. J., 136}, 280-290.

\refs Vokrouhlick\'y, D., and Nesvorn\'y, D. (2009) The common roots of
 asteroids (6070) Rheinland and (54827) 2001 NQ8. {\em Astron. J.,
 137}, 111-117.

\refs Vokrouhlick\'y, D., Milani, A., and Chesley, S. R. (2000) Yarkovsky
 effect on near-Earth asteroids: Mathematical formulation and
 examples. {\em Icarus, 148}, 118-138.

\refs Vokrouhlick\'y, D., Nesvorn\'y, D., and Bottke W. F. (2003) The
 vector alignments of asteroid spins by thermal torques. {\em Nature,
 425}, 147-151.

\refs Vokrouhlick\'y, D., Nesvorn\'y, D., and Bottke W. F. (2006d) Secular
 spin dynamics of inner main-belt asteroids. {\em Icarus, 184}, 1-28.

\refs Vokrouhlick\'y, D., Chesley, S. R., and Matson, R. D. (2008) Orbital
 identification for asteroid 152563 (1992 BF) through the
 Yarkovsky effect. {\em Astron. J., 135}, 2336-2340.

\refs Vokrouhlick\'y, D., \v{C}apek, D., Kaasalainen, M., and Ostro, S. J.
 (2004) Detectability of YORP rotational slowing of asteroid 25143 Itokawa.
 {\em Astron. Astrophys., 414}, L21-L24.

\refs Vokrouhlick\'y, D., \v{C}apek, D., Chesley, S. R., and Ostro, S. J.
 (2005a) Yarkovsky detection opportunities. I. Solitary asteroids. {\em
 Icarus, 173}, 166-184.

\refs Vokrouhlick\'y, D., \v{C}apek, D., Chesley, S. R., and Ostro, S. J.
 (2005b) Yarkovsky detection opportunities. II. Binary asteroids. {\em
 Icarus, 179}, 128-138.

\refs Vokrouhlick\'y, D., Breiter, S., Nesvorn\'y, D., and Bottke, W. F.
 (2007) Generalized YORP evolution: onset of tumbling and new
 asymptotic states. {\em Icarus, 191}, 636-650.

\refs Vokrouhlick\'y, D., Nesvorn\'y, D., Bottke, W. F., and Morbidelli, A.
 (2010) Collisionally born family about 87 Sylvia. {\em Astron. J., 139},
 2148-2158.

\refs Vokrouhlick\'y, D., Bro\v{z}, M., Bottke, W. F., Nesvorn\'y, D., and 
 Morbidelli, A. (2006a) Yarkovsky/YORP chronology of asteroid families.
 {\em Icarus, 182}, 118-142.

\refs Vokrouhlick\'y, D., Bro\v{z}, M., Bottke, W. F., Nesvorn\'y, D., and 
 Morbidelli, A. (2006b) The peculiar case of the Agnia asteroid family.
 {\em Icarus, 183}, 349-361.

\refs Vokrouhlick\'y, D., Bro\v{z}, M., Morbidelli, A., Bottke, W. F., Nesvorn\'y, D.,
 Lazzaro, D., and Rivkin, A. S. (2006c) Yarkovsky footprints in the Eos family.
 {\em Icarus, 182}, 92-117.

\refs Vokrouhlick\'y, D., \v{D}urech, J., Micha{\l}owski, T., et~al. (2009)
 Datura family: The 2009 update. {\em Astron. Astrophys., 507}, 495-504.

\refs Vokrouhlick\'y, D., \v{D}urech, J., Polishook, D., et~al. (2011) Spin
 vector and shape of (6070)~Rheinland and their implications. {\em Astron. J.,
 142}, 159(8pp).

\refs Vokrouhlick\'y, D., Farnocchia, D., \v{C}apek, D., et~al. (2015) The
 Yarkovsky effect for 99942 Apophis. {\em Icarus}, in press.

\refs Walsh, K. J., Richardson, D. C., and Michel, P. (2008) Rotational
 breakup as the origin of small binary asteroids. {\em Nature, 454}, 188-191.

\refs Yarkovsky, I. O. (1901) The density of luminiferous ether and the
 resistance it offers to motion. (in Russian) Bryansk, published privately by
 the author.

\refs \v{Z}i\v{z}ka, J., and Vokrouhlick\'y, D. (2011) Solar radiation
 pressure on (99942) Apophis. {\em Icarus, 211}, 511-518.

}

\end{document}